\documentclass[twocolumn]{revtex4}
\usepackage{amsmath,dcolumn,feynmf,epsfig,CJK}
\begin{document}
%\begin{CJK*}{GBK}{song}
\title{QUANTUM ELECTRODYNAMICS IN A LASER \\AND THE ELECTRON LASER
COLLISION}
\author{ZHANG QI-REN}
\address{School of Physics, Peking University, Beijing 100871,
China}

%\date{}
%\parskip 0pt
%\parindent 5mm
\begin{abstract}Quantum electrodynamics in a laser is formulated,
in which the electron-laser interaction is exactly considered, while
the interaction of an electron and a single photon is considered by
perturbation. The formulation is applied to the electron-laser
collisions. The effect of coherence between photons in the laser is
therefore fully considered in these collisions. The possibility of
$\gamma-$ray laser generation by use of this kind of collision is
discussed.


{\bf PACS number(s)}:12.20.-m, 11.80.-m, 42.55.Ah, 42.50.Ct,
42.55.Vc
\end{abstract} \maketitle
%\end{CJK*}

\vskip0.3cm

\section{Introduction}
The matter-laser interaction is an interesting topic in physics.
Many interesting new phenomena are waiting us to
explore\cite{v,a,b,g,z,le,ao,zz,x,r}. Theoretically, the problem is
how to solve the wave equation of charged particles under the joint
actions of the laser and of other origins. Therefore, the easiest
problem in this area is the electron-laser collision. To solve it
one needs only to solve the wave equation of an electron in a
classical electromagnetic field presenting the laser only. For a
plane wave laser it is already solved\cite{w,zq}. Of course, even in
a simple electron-laser collision, some photons other than those in
the source laser may be created or absorbed. The fundamental theory
for researching these processes has to be quantum electrodynamics.

Quantum electrodynamics is the most successful theory in
physics\cite{we}. Its main applications were to various processes
participated by a few photons and electrons only. In these
applications, it is called the quantum electrodynamics in vacuum.
Its results always agree with observations to a precision as high as
achievable by the present experiments. This is partially because the
smallness of the fundamental electromagnetic coupling constant, and
the renormalizability of the theory, so that the perturbation theory
is applicable. However, in the electron-laser processes, the
interaction, though is still of the electromagnetic origin, is not
weak because the coherence of huge number photons interacting with
the electron. To handle the non-perturbation character in these
processes one needs some new methods in quantum electrodynamics.
Here we formulate a quantum electrodynamics in laser, in which the
vacuum state of electromagnetic field in usual formulation is
substituted by a coherent state of photons describing the laser. In
the following we will show that, this formulation is a new picture
of quantum electrodynamics, equivalent to its other pictures, but
specially suitable for the treatment of electromagnetic processes
with a laser. The interaction of electrons with the laser is exactly
considered, but the interaction between electrons and the photons
other than those in the laser is still handled by perturbation. The
electron laser scattering is demonstrated as an example, with
special interest in the possible construction of a $\gamma -$ray
laser.

In section \ref{A2} we reform the quantum electrodynamics, so that
instead to quantize the electromagnetic field itself we quantize the
fluctuation of it around one of its classical solutions, the laser
field. As a contrast, the usual quantization method of
electrodynamics in vacuum is also shortly reviewed. In section
\ref{A3} we apply the obtained formulation to the case of quantum
electrodynamics in a circularly polarized laser, and transform it
into a picture, in which the Hamiltonian is divided into two parts,
one is time independent and another is a small perturbation. In
section \ref{A4} we solve the eigen-equation of the time independent
unperturbed Hamiltonian, obtain a complete orthonormal set of
eigenfunctions for the electron. By expanding the electron field in
terms of this set of eigenfunctions, we quantize the electron field
in the laser. The formulation is completed. In section \ref{A5} we
apply this formulation to the electron laser collision, the
expression of the cross section is obtained, and some examples of
numerical results are shown. In section \ref{A6} more numerical
results are shown. Especially, the case of large ${\cal N}$ is
discussed in some details. ${\cal N}$ is an integer, may be imagined
to be the number of photons in the laser, which participate the
collision. But from an analysis of the mathematical derivation we
see, its presence is a result of the electron wave distortion  by
the laser. In section \ref{A7} the possibility of making a
$\gamma-$ray laser by the electron laser collision is considered.
Section \ref{A8} is for discussions.

\section{Quantization of electromagnetic field around one of its
classical expression, the quantum electrodynamics in a
laser\label{A2}}

Consider an electrodynamic system composed of electrons and photons.
In the nature unit system ($c=\hbar=1$), its Lagrangian density is
\begin{equation}{\cal L}=-\bar{\Psi}\left[\gamma_\mu
(\partial_\mu+{\rm i}eA_\mu) + m\right]\Psi
-\frac{1}{4}F_{\mu\nu}F_{\mu\nu} ,\label{1}\end{equation} with
\begin{equation}F_{\mu\nu}=\partial_\mu A_\nu -\partial_\nu A_\mu .\label{2}\end{equation}
$\Psi$ denotes the electron field, [$A_\mu$] with $\mu=1,2,3,4$
denote the 4-electromagnetic potential, $e$ is the absolute value of
the electron charge, and $m$ is the electron mass. Notations in
Lurie's book\cite{l} are used here. By a standard procedure
collecting the contributions of longitudinal and temporal components
of the electromagnetic potential into a coulomb energy
\begin{equation}H_c=\frac{1}{2}\int {\rm d}^3x\int{\rm d}^3x'
\Psi^\dag(\mbox{\boldmath $x$})\Psi(\mbox{\boldmath $x$})
\frac{\alpha}{|\mbox{\boldmath $x$}-\mbox{\boldmath $x$}'|}
\Psi^\dag(\mbox{\boldmath $x$}')\Psi(\mbox{\boldmath
$x$}')\label{3}\end{equation} one obtains the Hamiltonian
\begin{equation}H=H_1+H_2+H_c \label{4}\end{equation}
of the system in Coulomb gauge, with
\begin{eqnarray}H_1&= &\int\Psi^\dag[ \mbox{\boldmath $\alpha$}\cdot
( \mbox{\boldmath $p$}+e \mbox{\boldmath $A$})+\beta m]\Psi{\rm d}^3x,\label{5}\\
H_2&=&\frac{1}{2}\int ({\cal E}^2+{\cal H}^2){\rm d}^3x.\label{6}
\end{eqnarray}
$\alpha\equiv e^2/4\pi=1/137.\cdots$ is the fine structure constant
showing the strength of the electromagnetic interaction;
$\mbox{\boldmath $p$}=-{\rm i}\mbox{\boldmath$\nabla$}$ is the
momentum of an electron; $\mbox{\boldmath ${\cal
E}$}=-\partial\mbox{\boldmath $A$}/\partial t$ is the transverse
electric field strength under the Coulomb gauge condition
\begin{equation}\mbox{\boldmath$\nabla$}\cdot\mbox{\boldmath $A$}=0;\label{6a}\end{equation}
$\mbox{\boldmath ${\cal
H}$}=\mbox{\boldmath$\nabla$}\times\mbox{\boldmath$A$}$ is the
magnetic field strength. Since $H_c$ is of the higher order of the
small parameter $\sqrt{\alpha}$, and is proportional to the square
of electron number density, one needs to consider the first two
terms only in eq.(\ref{4}) for processes participated by a few
electrons and photons, unless the electron number density in the
problem is as high as that in atoms. It is to consider the
Hamiltonian
\begin{equation}H_s=H_1+H_2. \label{7}\end{equation}

Expand $\Psi(\mbox{\boldmath $x$})$ and $\bar{\Psi}(\mbox{\boldmath
$x$})$ in terms of the complete orthonormal set of eigenfunctions
$\frac{1}{\sqrt{(2\pi)^3}}u_\sigma(\mbox{\boldmath
$p$})\exp\left({\rm i}\mbox{\boldmath $p$}\cdot\mbox{\boldmath
$x$}\right)$ and
$\frac{1}{\sqrt{(2\pi)^3}}v_{\sigma}(\mbox{\boldmath
$p$})\exp\left(-{\rm i}\mbox{\boldmath $p$}\cdot \mbox{\boldmath
$x$}\right)$ of the single electron energy operator $\mbox{\boldmath
$\alpha$}\cdot (-\mbox{i}\nabla)+\beta m$ in vacuum, in which
bispinors $u_{\sigma}(\mbox{\boldmath $p$})$ and
$v_{\sigma}(\mbox{\boldmath $p$})$ satisfy equations
\begin{eqnarray}&&\left(\mbox{\boldmath $\alpha$}\cdot \mbox{\boldmath $p$}+\beta
m\right)u_\sigma(\mbox{\boldmath $p$})=E(p) u_\sigma(\mbox{\boldmath
$p$})\, ,\\&&\left(\mbox{\boldmath $\alpha$}\cdot \mbox{\boldmath
$p$}-\beta m\right)v_\sigma(\mbox{\boldmath $p$})=E(p)
v_\sigma(\mbox{\boldmath $p$})\, ,\end{eqnarray} with
\begin{eqnarray}&&E(p)=\sqrt{p^2+m^2}\, ,\\&&u^\dag_\sigma(\mbox{\boldmath $p$})
u_{\sigma'}(\mbox{\boldmath $p$})=\delta_{\sigma\sigma'}\\
&&v^\dag_\sigma(\mbox{\boldmath $p$}) v_{\sigma'}(\mbox{\boldmath
$p$})= \delta_{\sigma\sigma'}\, ,\\&&u^\dag_\sigma(\mbox{\boldmath
$p$}) v_{\sigma'}(\mbox{\boldmath
$p$})=v^\dag_\sigma(\mbox{\boldmath $p$})
u_{\sigma'}(\mbox{\boldmath $p$})=0\, ,\end{eqnarray} $\sigma$ is
the spin index. We have
\begin{eqnarray}\Psi(\mbox{\boldmath $x$})&=&\int\frac{{\rm
d}^3p}{\sqrt{(2\pi)^3}}\!\sum_{\sigma}\!\left[c_{\sigma}(\mbox
{\boldmath $p$})u_{\sigma}(\mbox{\boldmath $p$})
 {\rm e}^{{\rm
i}\mbox {\boldmath $p$}\cdot\mbox {\boldmath
$x$}}\right.\nonumber\\&+&\left.d^\dag_{\sigma}(\mbox{\boldmath
$p$}) v_{\sigma}(\mbox {\boldmath $p$}){\rm e}^{-{\rm
i}\mbox{\boldmath $p$}\cdot\mbox
{\boldmath $x$}}\right]\, ,\label{8}\\
\bar\Psi(\mbox {\boldmath $x$})&=&\int\frac{{\rm
d}^3p}{\sqrt{(2\pi)^3}}\!\sum_{\sigma}\!\left[d_{\sigma}(\mbox
{\boldmath $p$})\bar{v}_{\sigma}(\mbox {\boldmath $p$}) {\rm
e}^{{\rm i}\mbox {\boldmath $p$}\cdot\mbox {\boldmath
$x$}}\right.\nonumber\\&+&\left.c^\dag_\sigma(\mbox {\boldmath
$p$})\bar {u}_\sigma(\mbox {\boldmath $p$}){\rm e}^{-{\rm i}\mbox
{\boldmath $p$}\cdot\mbox {\boldmath $x$}}\right]\,
.\label{9}\end{eqnarray} In the same way, expanding $\mbox
{\boldmath $A$}(\mbox{\boldmath $x$})$ in terms of transverse plane
waves $\frac{1}{\sqrt{(2\pi)^32k}}\,\mbox {\boldmath $e$}_i {\rm
e}^{{\rm i}\mbox {\boldmath $k$}\cdot\mbox {\boldmath $x$}},i=1,2,$
with
\begin{eqnarray}\mbox
{\boldmath $e$}^*_i\cdot\mbox
{\boldmath $e$}_{i'}&=&\delta_{ii'},\\
\mbox {\boldmath $e$}^*_i\cdot\mbox {\boldmath
$k$}&=&0,\end{eqnarray}we have\begin{eqnarray} \mbox {\boldmath
$A$}(\mbox {\boldmath $x$})\!\!=\!\!\!\int\!\!{\rm d}^3k\!\sum_i\!\!
\left[a_i(\mbox {\boldmath $k$})\frac{\mbox {\boldmath $e$}_i {\rm
e}^{{\rm i}\mbox {\boldmath $k$}\cdot\mbox {\boldmath
$x$}}}{\sqrt{(2\pi)^32k}}\!+\! a_i^\dag(\mbox {\boldmath
$k$})\!\frac{\mbox {\boldmath $e$}_i^* {\rm e}^{{-\rm i}\mbox
{\boldmath $k$}\cdot\mbox {\boldmath
$x$}}}{\sqrt{(2\pi)^32k}}\right]\!\!.\label{10}\end{eqnarray}
$c_\sigma,c_\sigma^\dag,d_\sigma,d_\sigma^\dag,a_i$ and $a_i^\dag$
are expansion coefficients. Quantization rules are\begin{eqnarray}
&& c_\sigma(\mbox {\boldmath $p$})c_{\sigma'}(\mbox {\boldmath
$p$}')
+c_{\sigma'}(\mbox {\boldmath $p$}')c_\sigma(\mbox {\boldmath $p$})\nonumber\\
&=&d_\sigma(\mbox {\boldmath $p$})d_{\sigma'}(\mbox {\boldmath $p$}')
+d_{\sigma'}(\mbox {\boldmath $p$}')d_\sigma(\mbox {\boldmath $p$})\nonumber\\
&=&c_\sigma(\mbox{\boldmath $p$})d_{\sigma'}(\mbox {\boldmath
$p$}')+d_{\sigma'}(\mbox {\boldmath $p$}')c_\sigma(\mbox {\boldmath
$p$})\nonumber\\&=&c_\sigma(\mbox {\boldmath
$p$})d^\dag_{\sigma'}(\mbox {\boldmath $p$}')+d^\dag_{\sigma'}(\mbox
{\boldmath $p$}')c_\sigma(\mbox {\boldmath $p$})=0\,
,\label{11}\\&&c_\sigma(\mbox{\boldmath $p$})c^\dag_{\sigma'}(\mbox {\boldmath $p$}')
+c^\dag_{\sigma'}(\mbox {\boldmath $p$}')c_\sigma(\mbox {\boldmath $p$})\nonumber\\
&=&d_\sigma(\mbox {\boldmath $p$})d^\dag_{\sigma'}(\mbox {\boldmath $p$}')
+d^\dag_{\sigma'}(\mbox {\boldmath $p$}')d_\sigma(\mbox {\boldmath $p$})\nonumber\\
&=&\delta_{\sigma\sigma'}\delta(\mbox {\boldmath $p$}-\mbox
{\boldmath $p$}')\, ,\label{12}\end{eqnarray}
\begin{eqnarray}
&&a_i(\mbox {\boldmath $k$})a_{i'}(\mbox {\boldmath
$k$}')-a_{i'}(\mbox {\boldmath $k$}')a_i(\mbox {\boldmath $k$})=0\,
, \label{13}\\&&a_i(\mbox {\boldmath $k$})a_{i'}^\dag(\mbox
{\boldmath $k$}')-a_{i'}^\dag(\mbox {\boldmath $k$}')a_i(\mbox
{\boldmath $k$}) =\delta_{ii'}\delta(\mbox {\boldmath $k$}-\mbox
{\boldmath $k$}')\, .\label{14}\end {eqnarray}$a_i(\mbox {\boldmath
$k$})$ and $a_i^\dag(\mbox {\boldmath $k$})$ always commute with
$c_\sigma(\mbox {\boldmath $p$})$ and $d_\sigma(\mbox {\boldmath
$p$})$. The vacuum state $|0\rangle$ is defined by
\begin{eqnarray}c_\sigma(\mbox{\boldmath $p$})|0\rangle
=d_\sigma(\mbox{\boldmath $p$})|0\rangle=a_i(\mbox {\boldmath
$k$})|0\rangle=0.\label{15}\end {eqnarray}It is a state without any
electron, positron or photon, and is therefore a vacuum state in its
usual meaning. This formulation is called the quantum
electrodynamics in vacuum.

In the Schr\"{o}dinger picture, operators $c_\sigma,d_\sigma$ and
$a_i$ are time independent, and the state vector $|t\rangle$ depends
on time $t$ according to the Schr\"{o}dinger equation
\begin{equation} {\rm i}\frac{{\rm d}|t\rangle}{{\rm
d}t}=H_s|t\rangle.\end{equation} Consider a time dependent unitary
transformation generated by the operator ${\rm e}^{{\rm i}H_2t}$. It
transforms the state vector $|\rangle$ and the operator $O$ into
\begin{eqnarray}|\rangle_{(ei)}\equiv{\rm e}^{{\rm
i}H_2t}|\rangle\;\;\;\mbox{and}\;\;\; O_{(ei)}\equiv{\rm e}^{{\rm
i}H_2t}O{\rm e}^{-{\rm i}H_2t}\label{c}\end{eqnarray}respectively.
Substituting (\ref{10}) into (\ref{6}) we have\begin{equation}
H_2=\int\sum_i\frac{k}{2}(a_i^\dag(\mbox{\boldmath
$k$})a_i(\mbox{\boldmath $k$})+a_i(\mbox{\boldmath
$k$})a_i^\dag(\mbox{\boldmath $k$})){\rm d}^3k.\end{equation} It
shows ${c_\sigma}_{(ei)}(\mbox{\boldmath
$p$})=c_\sigma(\mbox{\boldmath
$p$}),{d_\sigma}_{(ei)}(\mbox{\boldmath
$p$})=d_\sigma(\mbox{\boldmath $p$})$, but
${a_i}_{(ei)}(\mbox{\boldmath $k$})=a_i(\mbox{\boldmath $k$}){\rm
e}^{-{\rm i}kt}$. The free motion of the electromagnetic field is
therefore already considered in the transformation. In this picture,
the time dependence of the state is governed by the
equation\begin{equation} {\rm i}\frac{{\rm d}|t\rangle_{(ei)}}{{\rm
d}t}={H_1}_{(ei)}|t\rangle_{(ei)}.\end{equation}${H_1}_{(ei)}$ is
the Hamiltonian for the electron motion and the electron-photon
interaction in this picture, with\begin{eqnarray} \mbox{\boldmath
$A$}_{(ei)}(\mbox{\boldmath $x$})&=&\int{\rm d}^3k\sum_i
\left[a_i(\mbox{\boldmath $k$})\frac{\mbox{\boldmath $e$}_i {\rm
e}^{{\rm i}(\mbox{\boldmath $k$}\cdot\mbox{\boldmath
$x$}-kt)}}{\sqrt{(2\pi)^32k}}\right.\nonumber\\&+&\left.
a_i^\dag(\mbox{\boldmath $k$})\frac{\mbox{\boldmath $e$}_i^* {\rm
e}^{{-\rm i}(\mbox{\boldmath $k$}\cdot\mbox{\boldmath
$x$}-kt)}}{\sqrt{(2\pi)^32k}}\right].\end{eqnarray}  We may
therefore call this picture the electron-interaction picture, and
denote it by the subscript $(ei)$.

A laser is a classical limit of the intense electromagnetic wave and
is well described by the classical vector potential \begin{eqnarray}
\mbox{\boldmath $A$}_c(\mbox{\boldmath $x$})&=&\int{\rm d}^3k\sum_i
\left[a_{ic}(\mbox{\boldmath $k$})\frac{\mbox{\boldmath $e$}_i {\rm
e}^{{\rm i}(\mbox{\boldmath $k$}\cdot\mbox{\boldmath
$x$}-kt)}}{\sqrt{(2\pi)^32k}}\right.\nonumber\\&+&\left.
a_{ic}^*(\mbox{\boldmath $k$})\frac{\mbox{\boldmath $e$}_i^* {\rm
e}^{{-\rm i}(\mbox{\boldmath $k$}\cdot\mbox{\boldmath
$x$}-kt)}}{\sqrt{(2\pi)^32k}}\right],\label{b}
\end{eqnarray}$a_{ic}$ and $a_{ic}^*$ are c-number expansion coefficients.
Now, instead $\mbox{\boldmath $A$}(\mbox{\boldmath $x$})$,  let us
consider the fluctuation
\begin{eqnarray}
\mbox{\boldmath $A$}'(\mbox{\boldmath $x$})\!\!&\equiv&\!\!
\mbox{\boldmath $A$}_{(ei)}(\mbox{\boldmath
$x$})\!-\!\mbox{\boldmath $A$}_c(\mbox{\boldmath
$x$})\!\!=\!\!\int{\rm d}^3k\sum_i \left[a'_i(\mbox{\boldmath
$k$})\frac{\mbox{\boldmath $e$}_i {\rm e}^{{\rm i}(\mbox{\boldmath
$k$}\cdot\mbox{\boldmath
$x$}-kt)}}{\sqrt{(2\pi)^32k}}\right.\nonumber\\&+&\left.
{a'_i}^\dag(\mbox{\boldmath $k$})\frac{\mbox{\boldmath $e$}_i^* {\rm
e}^{{-\rm i}(\mbox{\boldmath $k$}\cdot\mbox{\boldmath
$x$}-kt)}}{\sqrt{(2\pi)^32k}}\right]\label{17}
\end{eqnarray}of vector potential $\mbox{\boldmath $A$}$ around the
laser $\mbox{\boldmath $A$}_c$, in which
\begin{eqnarray}
a'_i(\mbox{\boldmath $k$})=a_i(\mbox{\boldmath
$k$})-a_{ic}(\mbox{\boldmath $k$}).\label{a}
\end{eqnarray}Since $a_{ic}(\mbox{\boldmath $k$})$ and  $a_{ic}^*(\mbox{\boldmath $k$})$ are c-numbers, we see
\begin{eqnarray}
&&a'_i(\mbox{\boldmath $k$})a'_{i'}(\mbox{\boldmath
$k$}')-a'_{i'}(\mbox{\boldmath $k$}')a'_i(\mbox{\boldmath $k$})=0\,
,\\&&a'_i(\mbox{\boldmath $k$}){a'_{i'}}^\dag(\mbox{\boldmath
$k$}')-{a'_{i'}}^\dag(\mbox{\boldmath $k$}')a'_i(\mbox{\boldmath
$k$}) =\delta_{ii'}\delta(\mbox{\boldmath $k$}-\mbox{\boldmath
$k$}')\, .
\end{eqnarray}
These equations may also be viewed as the quantization rules for the
electromagnetic field, equivalent to equations (\ref{13}) and
(\ref{14}). This is the quantization of electromagnetic field around
one of its classical process. The 'vacuum state'
$|[a_{i'c}(\mbox{\boldmath $k$}')];0\rangle$ of electromagnetic
field fluctuation in this formulation is defined by
\begin{eqnarray}a'_i(\mbox{\boldmath $k$})|[a_{i'c}(\mbox{\boldmath $k$}')];0\rangle=0.\label{r}\end{eqnarray}
By (\ref{a}), we see
\begin{eqnarray}a_i(\mbox{\boldmath $k$})|[a_{i'c}(\mbox{\boldmath $k$}')];0\rangle=
a_{ic}(\mbox{\boldmath $k$})|[a_{i'c}(\mbox{\boldmath
$k$}')];0\rangle.\end{eqnarray} It shows that this is a coherent
state of electromagnetic field with amplitudes
$[a_{i'c}(\mbox{\boldmath $k$}')]$, describing the laser (\ref{b}).
We call this formulation the quantum electrodynamics in a laser.
\section{Quantum electrodynamics in a circularly polarized laser, the
rotation picture\label{A3}}

In the Coulomb gauge, a circularly polarized laser is well described
by the classical vector potential
\begin{eqnarray}
\mbox{\boldmath $A$}_c(\mbox{\boldmath $x$})=A_c\{\mbox{\boldmath
$x$}_0\cos[k_c(z-t)] +\mbox{\boldmath $y$}_0 \sin[k_c(z-t)]\}\
.\label{16}
\end{eqnarray}
It is a plane wave circularly polarized in the $x-y$ plane and
propagating along the $z$ direction, with a wave vector
$\mbox{\boldmath $k$}_c=k_c\mbox{\boldmath $z$}_0$ and an amplitude
$A_c$. $\mbox{\boldmath $x$}_0$, $\mbox{\boldmath $y$}_0$ and
$\mbox{\boldmath $z$}_0$ are unit vectors along $x$, $y$ and $z$
directions respectively. Writing it in the form (\ref{b}), we see
\begin{eqnarray}
a_{ic}(\mbox{\boldmath $k$})=\sqrt{(2\pi)^3k_c}\;A_c\delta_
{i1}\delta(\mbox{\boldmath $k$}-\mbox{\boldmath
$k$}_c)\end{eqnarray}and\begin{eqnarray}\mbox{\boldmath
$e$}_1\equiv\mbox{\boldmath $e$}=\frac{\mbox{\boldmath $x$}_0-{\rm
i}\mbox{\boldmath $y$}_0}{\sqrt{2}}\;\;\;\;\;\;(\mbox{for}\;
\mbox{\boldmath $k$}=\mbox{\boldmath $k$}_c).
\end{eqnarray}
By equations (\ref{5}),(\ref{c}) and (\ref{17}) we obtain
\begin{eqnarray}{H_1}_{(ei)}&=&H_0+H',\\
H_0&=& \int\Psi^\dag[ \mbox{\boldmath $\alpha$}\cdot
( \mbox{\boldmath $p$}+e \mbox{\boldmath $A$}_c)+\beta m]\Psi{\rm d}^3x,\\
H'&=&e\int\Psi^\dag\mbox{\boldmath $\alpha$}\cdot  \mbox{\boldmath
$A$}'\Psi{\rm d}^3x.
\end{eqnarray}Equation (\ref{16}) shows, $H_0$ is time dependent.
Fortunately, a time dependent unitary transformation generated by
the operator ${\rm e}^{-{\rm i}k_ctJ_z}$ may remove the time
dependence of $H_0$, in which\begin{equation} J_z\equiv\int\Psi^\dag
j_z\Psi{\rm d}^3x\end{equation} is the $z$-component of the angular
momentum for the electron system, with
\begin{equation}j_z=-{\rm i}\frac{\partial}{\partial
\varphi}+\frac{\Sigma_z}{2}\end{equation} being the $z$-component of
the angular momentum for one electron. $\varphi$ is the azimuth
angle of the electron and $\Sigma_z$ is the $z$-component of the
Pauli matrices. This transformation changes the state vector
$|\rangle$ and operator $O$ into\begin{equation} |\rangle_r={\rm
e}^{-{\rm i}k_ctJ_z}|\rangle \;\;\;\mbox{and}\;\;\;O_r={\rm
e}^{-{\rm i}k_ctJ_z}O{\rm e}^{{\rm
i}k_ctJ_z}\label{s0}\end{equation} respectively. It is a rotation
around the $z$-direction with angular velocity $k_c$, the resulting
picture is therefore called the rotation picture, and denoted by the
subscript $r$. Simple derivation shows
\begin{equation} {H_0}_r\!\!=\!\!\!\int\!\!\Psi^\dag\!\{ \mbox{\boldmath $\alpha$}\cdot
\mbox{\boldmath $p$}+eA_c[
\alpha_x\!\cos(k_cz\!)+\alpha_y\!\sin(k_cz\!)]\!+\!\beta m\}\Psi{\rm
d}^3\!x,\end{equation} which is indeed time independent. The time
dependence of state vector in this picture is governed
by\begin{equation} {\rm i}\frac{\partial |\rangle_r }{\partial
t}=(k_cJ_z+{H_0}_r+{H'}_r)|\rangle_r .\label{d}\end{equation}Take
\begin{eqnarray}{H_0}_{r+}&\equiv&{H_0}_r+k_cJ_z=\int\Psi^\dag\{
\mbox{\boldmath $\alpha$}\cdot \mbox{\boldmath $p$}+eA_c[
\alpha_x\cos(k_cz)\nonumber\\&+&\alpha_y\sin(k_cz)]+\beta
m+k_cj_z\}\Psi{\rm d}^3x\end{eqnarray}to be the unperturbed
Hamiltonian, and
\begin{eqnarray}&&{H'}_r=e\int{\rm d}^3x\int{\rm d}^3k\sum_i
\Psi^\dag\left[a'_i(\mbox{\boldmath $k$})\frac{\mbox{\boldmath
$\alpha$}\cdot\mbox{\boldmath $e$}_{ir} {\rm e}^{{\rm
i}(\mbox{\boldmath $k$}_r\!\cdot\,\mbox{\boldmath
$x$}-kt)}}{\sqrt{(2\pi)^32k}}\right.\nonumber\\&&\left.+{a'_i}^\dag(\mbox{\boldmath
$k$})\frac{\mbox{\boldmath $\alpha$}\cdot\mbox{\boldmath $e$}_{ir}^*
{\rm e}^{{-\rm i}(\mbox{\boldmath $k$}_r\!\cdot\,\mbox{\boldmath
$x$}-kt)}}{\sqrt{(2\pi)^32k}}\right]\Psi\label{t0}\end{eqnarray} to
be the perturbation, with\begin{eqnarray}\mbox{\boldmath
$k$}_r&=&[k_x\cos(k_ct)-k_y\sin(k_ct)]\mbox{\boldmath
$x$}_0\nonumber\\&+&
[k_x\sin(k_ct)+k_y\cos(k_ct)]\mbox{\boldmath $y$}_0+k_z\mbox{\boldmath $z$}_0,\label{t1}\\
\mbox{\boldmath
$e$}_{ir}&=&[e_{ix}\cos(k_ct)-e_{iy}\sin(k_ct)]\mbox{\boldmath
$x$}_0\nonumber\\&+&
[e_{ix}\sin(k_ct)+e_{iy}\cos(k_ct)]\mbox{\boldmath
$y$}_0+e_{iz}\mbox{\boldmath $z$}_0,\label{t2}\end{eqnarray} one may
solve the equation (\ref{d}) by the usual perturbation procedure.
The smallness of electromagnetic coupling constant $\sqrt{\alpha}$
makes the procedure reliable.

From eqs. (\ref{t1}) and (\ref{t2}) we see that plane waves
$\mbox{\boldmath$e$}_{ir} {\rm e}^{{\rm i}(\mbox{\boldmath
$k$}_r\!\cdot\,\mbox{\boldmath $x$}-kt)}$ in the interaction
Hamiltonian (\ref{t0}) rotate around the z axis with angular
velocity $k_c$. The picture transformation (\ref{s0}) stops the
rotation of the circularly polarized laser environment in $H_0$ and
starts the rotation of the quantum modes of the electromagnetic
field fluctuations around an opposite direction in $H'$. However,
the contents of the theory and the resulting prediction of
observation do not change.
\section{Quantization of the electron field in a circularly
polarized laser and the rotation-interaction picture\label{A4}}

To quantize the electron field $\Psi$ in the rotation picture, one
has to expand it in terms of a set of orthonormal functions
$[\psi_n]$, which diagonalizes ${H_0}_{r+}$. It
requires\begin{eqnarray}&\{&\!\!\!\!\mbox{\boldmath
$\alpha$}\!\cdot\!(-{\rm
i}\mbox{\boldmath$\nabla$})\!+\!eA_c[\alpha_x\cos
(k_cz)\!+\!\alpha_y\sin (k_cz)]\nonumber\\&+&\beta
m\!+\!k_cj_z\}\psi_n(\mbox{\boldmath$x$})=\varepsilon_n\psi_n(\mbox{\boldmath$x$}).\label{l}\end{eqnarray}
This is an eigenequation,  $\psi_n$ is the eigenfunction and $
\varepsilon_n$ is the eigenvalue. We derive the eigenfunctions  from
the Wolkow solution\cite{w,zq} of the Dirac equation for an electron
in a plane electromagnetic wave. In the circularly polarized plane
wave (\ref{16}), the Dirac equation is\begin{equation} {\rm
i}\frac{\partial\psi}{\partial t}=\{\mbox{\boldmath
$\alpha$}\cdot(-{\rm i}\mbox{\boldmath$\nabla$})+eA_c[\alpha_x\cos
\phi+\alpha_y\sin \phi]+\beta m\}\psi,\label{h}\end{equation} with
$\phi\equiv k_c(z-t)$. Its Wolkow solution is
\begin{equation}\psi_w(x)={\rm e}^{{\rm i}p_\mu x_\mu}{\cal F}(\phi),\end{equation}
with
\begin{eqnarray}&&{\cal F}(\phi)=\{1-eA_ck_c
[\alpha_x\cos \phi+\alpha_y\sin \phi\nonumber\\&+&\!\!\!{\rm
i}(\Sigma_y\cos \phi-\Sigma_x\sin \phi)]/2k_\mu p_\mu\}\exp\{-{\rm
i}e[2A_c(p_x\sin\phi\nonumber\\&-&\!\!\!p_y\cos\phi)+eA_c^2\phi+C]/2
k_\mu p_\mu\}u,
\end{eqnarray}
$C$ is a constant, $u$ is a Dirac bispinor satisfying the Dirac
equation
\begin{equation}(\gamma_\mu p_\mu-{\rm i}m)u=0
\label{e}\end{equation} for a free electron.The solution is
characterized by quantum numbers $p_\mu,\mu=1,2,3,4,$ satisfying the
energy-momentum relation $p_\mu p_\mu=-m^2$ of a free electron.
Write $p_x=p_{_\perp}\cos\varphi_p$ and
$p_y=p_{_\perp}\sin\varphi_p,$ we see\begin{widetext}
\begin{eqnarray}\psi_r(\mbox{\boldmath$x$},t)&\equiv&{\rm
e}^{-{\rm i}k_ctJ_z}\psi_w(\mbox{\boldmath$x$},t)={\rm e}^{{\rm
i}(p_zz-Et)}\left\{1-\frac{eA_c}{2(p_z-E)}\left[\alpha_x\cos k_cz+
\alpha_y\sin k_cz+{\rm i}(\Sigma_y\cos k_c z-\Sigma_x\sin
k_cz)\right]\right\}\nonumber\\&\times&\!\!\!\exp\left\{{\rm
i}p_{_\perp} \left[(x-\frac{eA_c}{k_c(p_z-E)}\sin k_cz)\cos
(k_ct+\varphi_p)+(y+\frac{eA_c}{k_c(p_z-E)}\cos
k_cz)\sin(k_ct+\varphi_p)\right]\right\}\nonumber\\&\times&\!\!\!\exp\left[-{\rm
i}\frac{e^2A_c^2(z-t)}{2(p_z-E)}-{\rm i}C'\right]{\rm e}^{-{\rm
i}\frac{k_ct\Sigma_z}{2}}u.\label{f}\end{eqnarray}\end{widetext}$C'$
is another constant, $E=\pm\sqrt{\mbox{\boldmath $p$}^2+m^2}$ is the
energy of a free electron with momentum \mbox{\boldmath $p$}.
Solving (\ref{e}) one has\begin{eqnarray}u=u_++u_-{\rm e}^{{\rm
i}\sigma\varphi\!_p},\label{v6}\end{eqnarray}\begin{eqnarray}
u_+\!\!=\!\!\sqrt{\frac{E+m}{2E}}\!\left[\!\!\begin{array}{c}1\\
\frac{p_z\sigma}{E+m}\end{array}\!\!\right]\!\!
\chi_{_\sigma},\;\;u_-\!\!=\!\!\sqrt{\frac{E+m}{2E}}\!\left[\!\!\begin{array}{c}0\\
\frac{p_\perp}{E+m}\end{array}\!\!\right]\!\!\chi\!_{_{-\!\sigma}},\label{u}
\end{eqnarray}
$\sigma=\pm 1$, and $\chi_{_{\pm\sigma}}$ is an eigen-spinor of
$\Sigma_z$ with eigenvalue $\pm \sigma$.Write
\begin{eqnarray}x=\rho\cos\varphi,\;\;\;\;\;y=\rho\sin\varphi,\end{eqnarray}
\begin{eqnarray}\left.\begin{array}{c}x'=x-\frac{eA_c}{k_c(p_z-E)} \sin
k_cz=\rho'\cos\varphi',\\y'=y+\frac{eA_c}{k_c(p_z-E)} \cos
k_cz=\rho'\sin\varphi'\end{array}\right\}.\end{eqnarray}They define
$\rho,\varphi,x',y',\rho'$ and $\varphi'$. It is a $z$-dependent
coordinate transformation from $x,y$ to $\rho',\varphi'$.Using the
formula
\begin{equation}{\rm e}^{{\rm i}a\cos\theta}=\sum_{n=-\infty}^\infty{\rm i}^n{\rm J}_n(a)
{\rm e}^{{\rm i}n\theta},\label{v2} \end{equation}in which ${\rm
J}_n(a)$ is a Bessel function of order $n$ in variable $a$, we find
\begin{widetext}\begin{eqnarray}\psi_r(\mbox{\boldmath$x$},t)&=&{\rm
e}^{{\rm
i}(p_zz-Et)}\left\{1-\frac{eA_c}{2(p_z-E)}\left[\alpha_x\cos k_cz+
\alpha_y\sin k_cz+{\rm i}(\Sigma_y\cos k_c z-\Sigma_x\sin
k_cz)\right]\right\}\nonumber\\&\times&\exp\left\{{\rm
i}\left[p_{_\perp}\rho'\cos(k_ct+\varphi_p-\varphi')
-\frac{e^2A_c^2(z-t)}{2(p_z-E)}-C'\right]\right\} [{\rm e}^{-{\rm
i}\frac{\sigma}{2}k_ct}u_++{\rm e}^{{\rm
i}\frac{\sigma}{2}(k_ct+2\varphi\!_p)}u_-]\nonumber\\
&=&{\rm e}^{{\rm
i}(p_zz-Et+\frac{\sigma}{2}\varphi\!_p)}\left\{1-\frac{eA_c}{2(p_z-E)}\left[\alpha_x\cos
k_cz+ \alpha_y\sin k_cz+{\rm i}(\Sigma_y\cos k_c z-\Sigma_x\sin
k_cz)\right]\right\}\nonumber\\&\!\!\!\!\!\times\!\!\!\!\!&\exp\left\{-{\rm
i}\left[\frac{e^2A_c^2(z-t)}{2(p_z-E)}+C'\right]\right\}\!\!\sum_{n=-\infty}^\infty\!\!\!\!{\rm
i}^n{\rm J}_n(p_{_\perp}\rho'){\rm e}^{-{\rm i}n\varphi'}[{\rm
e}^{{\rm i}(n-\frac{\sigma}{2})(k_ct+\varphi\!_p)}u_+\!\!+\!{\rm
e}^{{\rm
i}(n+\frac{\sigma}{2})(k_ct+\varphi\!_p)}u_-]\nonumber\\&=&\sum_{n=-\infty}^\infty{\rm
e}^{{\rm i}(n\varphi\!_p-C')}{\rm e}^{-{\rm i}\varepsilon_n
t}U_n(\mbox{\boldmath$x$}),\label{g}\end{eqnarray}with
\begin{equation}\varepsilon_n =E+\frac{e^2A_c^2}{2(E-p_z)}
+\left(\frac{\sigma}{2}-n\right)k_c,\label{i}\end{equation}
\begin{eqnarray}U_n(\mbox{\boldmath$x$})&=&\exp\left\{{\rm
i}\left[p_z+\frac{e^2A_c^2}{2(E-p_z)}\right]z\right\}\nonumber
\\&\times&\left\{1-\frac{eA_c}{2(p_z-E)}\left[\alpha_x\cos
k_cz+ \alpha_y\sin k_cz+{\rm i}(\Sigma_y\cos k_c z-\Sigma_x\sin
k_cz)\right]\right\}\nonumber\\&\times&\left[{\rm i}^n{\rm
J}_n(p_{_\perp}\rho'){\rm e}^{-{\rm i}n\varphi'}{\rm P}_++{\rm
i}^{n-\sigma}{\rm J}_{n-\sigma}(p_{_\perp}\rho'){\rm e}^{-{\rm
i}(n-\sigma)\varphi'}{\rm
P}_-\right]u_\sigma(0),\label{k}\end{eqnarray}\end{widetext} in
which ${\rm P}_\pm=\frac{1\pm\sigma\Sigma_z}{2}$ is the projection
operator, $u_\sigma(0)=u_++u_-$ is $u$ for $\varphi_p=0$. Since
$\psi_w$ is a solution of the Dirac equation (\ref{h}), $\psi_r$
defined in (\ref{f}) is a solution of the equation
\begin{eqnarray}{\rm i}\frac{\partial\psi}{\partial
t}&=&\{\mbox{\boldmath $\alpha$}\cdot(-{\rm
i}\mbox{\boldmath$\nabla$})+eA_c[\alpha_x\cos (k_cz)+\alpha_y\sin
(k_cz)]\nonumber\\&+&\beta m+k_cj_z\}\psi\,.\label{v4}\end{eqnarray}
Substituting the last expression of the equation (\ref{g}) into this
equation, we
see\begin{eqnarray}&&\sum_{n=-\infty}^\infty\!\!\!\!{\rm e}^{{\rm
i}(n\varphi\!_p-C')}{\rm e}^{-{\rm i}\varepsilon_n
t}\{\mbox{\boldmath $\alpha$}\!\cdot\!(-{\rm
i}\mbox{\boldmath$\nabla$})\!+\!eA_c[\alpha_x\cos
(k_cz)\nonumber\\&&+\alpha_y\sin (k_cz)]\!+\!\beta
m\!+\!k_cj_z-\varepsilon_n\}U_n(\mbox{\boldmath$x$})\!\!=\!\!
0.\label{j}\end{eqnarray}For given {\boldmath$p$}, $E$ and $\sigma$,
$\varepsilon_n$ is a monotonic function of $n$. Functions ${\rm
e}^{-{\rm i}\varepsilon_n t}$ are linearly independent of each other
for different $n$. The necessary and sufficient condition for the
validity of equation (\ref{j}) is\begin{eqnarray}&&\{\mbox{\boldmath
$\alpha$}\!\cdot\!(-{\rm
i}\mbox{\boldmath$\nabla$})\!+\!eA_c[\alpha_x\cos
(k_cz)\!+\!\alpha_y\sin (k_cz)]\nonumber\\&&+\!\beta
m\!+\!k_cj_z\}U_n(\mbox{\boldmath$x$})=\varepsilon_nU_n(\mbox{\boldmath$x$}).
\label{p}\end{eqnarray} Comparing with equation (\ref{l}) we see
$\psi_n(\mbox{\boldmath$x$})=N_nU_n(\mbox{\boldmath$x$})$, $N_n$ is
the normalization constant. Equation (\ref{k}) shows
$\psi_n(\mbox{\boldmath$x$})$ is characterized by five quantum
numbers, besides $n$ they are $p_z, p_{_\perp}, \sigma$, and $\tau$.
$\tau=\pm 1$ is defined by $E=\tau\sqrt{\mbox{\boldmath$p$}^2+m^2}$.
To save writing, we keep only one subscript $n$ to be the
representative of these five quantum numbers. To find the
normalization constant, we have to complete the integration
\begin{equation}I=\int
U^\dag_n(\mbox{\boldmath$x$})U_{n'}(\mbox{\boldmath$x$}){\rm
d}^3x.\end{equation}Since $U_n(\mbox{\boldmath$x$})$ is an
eigenfunction of an Hermite operator with eigenvalue
$\varepsilon_n$, $I$ is nonzero only when
$\varepsilon_n=\varepsilon_{n'}$.

Eq. (\ref{k}) shows that the coordinates appeared in two factors of
the integrand above are different. To complete the integration we
have to choose one common set of coordinates in both factors. Graf
formula\cite{wa,wg} solves this problem. Take the cylindric
coordinates $\rho,\varphi, z$ for {\boldmath$x$} in both factors,
and use the formula
\begin{eqnarray}&&{\rm J}_n(p_{_\perp}\rho'){\rm e}^{-{\rm
i}n\varphi'}\nonumber\\&&=\!\!\!\!\sum_{\nu=-\infty}^\infty\!\!\!\!{\rm
J}_{n-\nu}(p_{_\perp}R){\rm J}_\nu(p_{_\perp}\rho){\rm e}^{-{\rm
i}\nu\varphi}{\rm e}^{-{\rm
i}(n-\nu)(k_cz-\frac{\pi}{2})},\label{m}\end{eqnarray}in which
$R=\frac{eA_c}{k_c(E-p_z)}$ is of the length dimension. We also
define $R'=\frac{eA_c}{k_c(E'-p'\!_z)}$ for following derivations.
The remaining calculation becomes elementary, though is still
tedious. Main steps are shown in the appendix. The result is
\begin{equation}I=\frac{4\pi^2}{p_\perp}\delta(p_{_\perp}-p'_{_\perp})
\delta(p_z-p'_z)\delta_{n,n'}\delta_{\sigma,\sigma'}\delta_{\tau,\tau'}.\label{x}\end{equation}It
is simple and nice. The normalization constant $N_n=\frac{1}{2\pi}$
makes eigenfunctions
$\psi_n(\mbox{\boldmath$x$})=\frac{1}{2\pi}U_n(\mbox{\boldmath$x$})$
satisfy the orthonormal relations
\begin{equation}\int\!\!\psi^\dag_n(\mbox{\boldmath$x$})\psi_{n'}(\mbox{\boldmath$x$}){\rm d}^3x
\!=\!\frac{1}{p_{_\perp}}\delta(p_{_\perp}-p'_{_\perp})
\delta(p_z-p'_z)\delta_{n,n'}\delta_{\sigma,\sigma'}\delta_{\tau,\tau'}.\label{q}\end{equation}
A careful analysis further shows, that the orthonormal set
$[\psi_n(\mbox{\boldmath$x$})]$ is complete.

Expand the electron field function $\Psi(\mbox{\boldmath$x$})$ in
terms of $[\psi_n(\mbox{\boldmath$x$})]$. Writing
$\psi_n(\mbox{\boldmath$x$})$ in the form
$\psi_{n,\sigma,\tau}(p_z,p_{_\perp};\mbox{\boldmath$x$})$, defining
\begin{eqnarray}\left.\begin{array}{l}U_{n,\sigma}(p_z,p_{_\perp};\mbox{\boldmath$x$})\equiv
\psi_{n,\sigma,1}(p_z,p_{_\perp};\mbox{\boldmath$x$})\\
V_{n,\sigma}(p_z,p_{_\perp};\mbox{\boldmath$x$})\equiv
\psi_{-n,-\sigma,-1}(-p_z,p_{_\perp};\mbox{\boldmath$x$})\end{array}\right\},\label{v5}\end{eqnarray}
we have
\begin{eqnarray}\Psi(\mbox{\boldmath$x$})\!\!\!\!&=&\!\!\!\!\!\!\int_{-\infty}^\infty\!\!\!\!\!\!\!{\rm
d}p_z\!\!\int_0^\infty\!\!\!\!\!\!\! p_{_\perp}\!{\rm
d}p_{_\perp}\!\!\!\!\!\sum_{n=-\infty}^\infty \sum_{\sigma=\pm 1}\![
c_{n,\sigma}(p_z,p_{_\perp})U_{n,\sigma}(p_z,p_{_\perp};\mbox{\boldmath$x$})\nonumber\\
&+&d^\dag_{n,\sigma}(p_z,p_{_\perp})V_{n,\sigma}(p_z,p_{_\perp};\mbox{\boldmath$x$})].
\end{eqnarray}Using the orthonomal relations (\ref{q}) we obtain
expansion coefficients
\begin{eqnarray}\left.\begin{array}{l}c_{n,\sigma}(p_z,p_{_\perp})\!\!=\!\!\int U^\dag_{n,\sigma}
(p_z,p_{_\perp};\mbox{\boldmath$x$})\Psi(\mbox{\boldmath$x$}){\rm
d}^3x\\d^\dag_{n,\sigma}(p_z,p_{_\perp})\!\!=\!\!\int
V^\dag_{n,\sigma}
(p_z,p_{_\perp};\mbox{\boldmath$x$})\Psi(\mbox{\boldmath$x$}){\rm
d}^3x\end{array}\right\}.\end{eqnarray}The quantization rules
(\ref{11},\ref{12}) may be equivalently written in the
form\begin{eqnarray}\left.\begin{array}{l}\Psi_\iota(\mbox{\boldmath$x$})\Psi_{\iota'}(\mbox{\boldmath$x$}')
+\Psi_{\iota'}(\mbox{\boldmath$x$}')\Psi_\iota(\mbox{\boldmath$x$})=0\\
\Psi_\iota(\mbox{\boldmath$x$})\Psi^\dag_{\iota'}(\mbox{\boldmath$x$}')
+\Psi^\dag_{\iota'}(\mbox{\boldmath$x$}')\Psi_\iota(\mbox{\boldmath$x$})
=\delta_{\iota,\iota'}\delta(\mbox{\boldmath$x$}-\mbox{\boldmath$x$}')\end{array}\right\},\end{eqnarray}
$\iota$ and $\iota'=1,2,3,4$, are component indexes, which in turn
means\begin{eqnarray}&&c_{n,\sigma}(p_z,p_{_\perp})c_{n',\sigma'}(p'_z,p_{_\perp}')
+c_{n',\sigma'}(p'_z,p_{_\perp}')c_{n,\sigma}(p_z,p_{_\perp})\nonumber\\
&=&d_{n,\sigma}(p_z,p_{_\perp})d_{n',\sigma'}(p'_z,p_{_\perp}')
+d_{n',\sigma'}(p'_z,p_{_\perp}')d_{n,\sigma}(p_z,p_{_\perp})\nonumber\\
&=&c_{n,\sigma}(p_z,p_{_\perp})d_{n',\sigma'}(p'_z,p_{_\perp}')+
d_{n',\sigma'}(p'_z,p_{_\perp}')c_{n,\sigma}(p_z,p_{_\perp})\nonumber\\
&=&c_{n,\sigma}(p_z,p_{_\perp})d^\dag_{n',\sigma'}(p'_z,p_{_\perp}')
+d^\dag_{n',\sigma'}(p'_z,p_{_\perp}')c_{n,\sigma}(p_z,p_{_\perp})\nonumber\\&=&0\,
,\\&&c_{n,\sigma}(p_z,p_{_\perp})c^\dag_{n',\sigma'}(p'_z,p_{_\perp}')
+c^\dag_{n',\sigma'}(p'_z,p_{_\perp}')c_{n,\sigma}(p_z,p_{_\perp})\nonumber\\
&=&d_{n,\sigma}(p_z,p_{_\perp})d^\dag_{n',\sigma'}(p'_z,p_{_\perp}')
+d^\dag_{n',\sigma'}(p'_z,p_{_\perp}')d_{n,\sigma}(p_z,p_{_\perp})\nonumber\\
&=&\frac{1}{p_{_\perp}}\delta(p_{_\perp}-p_{_\perp}')\delta(p_z-p_z')\delta_{n,n'}\delta_{\sigma,\sigma'}\,
.\end{eqnarray}They are quantization rules for the electron field in
a laser. The 'vacuum state' $|[a_{i'c}(\mbox{\boldmath
$k$}')];0\rangle$ in this formulation is therefore defined, besides
(\ref{r}),
by\begin{eqnarray}\left.\begin{array}{l}c_{n,\sigma}(p_z,p_{_\perp})|[a_{i'c}(\mbox{\boldmath
$k$}')];0\rangle=0\\d_{n,\sigma}(p_z,p_{_\perp})|[a_{i'c}(\mbox{\boldmath
$k$}')];0\rangle=0\end{array}\right\},\label{t}\end{eqnarray}showing
that the numbers of electron and positron in vacuum states are
zeros. Writing $\varepsilon_n$ in the form
$\varepsilon_{n,\sigma,\tau}(p_z,p_{_\perp})$, taking the normal
product, we see the unperturbed Hamiltonian
\begin{eqnarray}&&\!\!\!\!{H_0}_{r+}=\nonumber\\&&\!\!\!\!\int_{-\infty}^\infty\!\!\!\!\!\!\!\!{\rm d}p_z
\!\!\int_0^\infty\!\!\!\!\!\!\! p_{_\perp}{\rm
d}p_{_\perp}\!\!\!\!\!\!\sum_{n=-\infty}^\infty\!\sum_{\sigma=\pm1}\!\!\!
[\varepsilon_{n,\sigma,1}\!(p_z,p_{_\perp}\!)c^\dag_{n,\sigma}
\!(p_z,\!p_{_\perp}\!)c_{n,\sigma}\!(p_z,\!p_{_\perp}\!)\nonumber\\
&&\!\!\!\!-\varepsilon_{-n,-\sigma,-1}(-p_z,\!p_{_\perp}\!)d^\dag_{n,\sigma}
\!(p_z,\!p_{_\perp}\!)d_{n,\sigma}\!(p_z,\!p_{_\perp}\!)].\end{eqnarray}

Consider, once again, a time dependent unitary transformation
generated by the operator ${\rm e}^{{\rm i}{H_0}_{r+}t}$. It
transforms the state vector $|\rangle$ and the operator $O$ into
\begin{equation}|\rangle_{(ri)}={\rm e}^{{\rm i}{H_0}_{r+}t}|\rangle
\;\;\;\;\mbox{and}\;\;\;\;O_{(ri)}={\rm e}^{{\rm i}{H_0}_{r+}t}O{\rm
e}^{-{\rm i}{H_0}_{r+}t}\end{equation}respectively. The time
dependence of the state vector now is governed by the
equation\begin{equation}{\rm i}\frac{{\rm d}|t\rangle_{(ri)}}{{\rm
d}t}=H'_{(ri)}|t\rangle_{(ri)},\end{equation}in
which\begin{eqnarray}&&H'_{(ri)}\!\!=e\!\!\int\!\!{\rm
d}^3x\!\!\int\!{\rm d}^3k\!\sum_i
\Psi_{(ri)}^\dag\!\!\left[a'_i(\mbox{\boldmath
$k$})\frac{\mbox{\boldmath $\alpha$}\cdot\mbox{\boldmath $e$}_{ir}
{\rm e}^{{\rm i}(\mbox{\boldmath $k$}_r\!\cdot\,\mbox{\boldmath
$x$}-kt)}}{\sqrt{(2\pi)^32k}}\right.\nonumber\\&&\left.+
{a'_i}^\dag(\mbox{\boldmath $k$})\frac{\mbox{\boldmath
$\alpha$}\cdot\mbox{\boldmath $e$}_{ir}^* {\rm e}^{{-\rm
i}(\mbox{\boldmath $k$}_r\!\cdot\,\mbox{\boldmath
$x$}-kt)}}{\sqrt{(2\pi)^32k}}\right]\Psi_{(ri)}\label{s}\end{eqnarray}is
the interaction Hamiltonian in this picture,
with\begin{eqnarray}&&\!\!\!\!\Psi_{(ri)}(\mbox{\boldmath$x$},t)=\!\!\int_{-\infty}^\infty\!\!\!\!\!\!\!\!
{\rm d}p_z\!\!\int_0^\infty\!\!\!\!\!\!\! p_{_\perp}{\rm
d}p_{_\perp}\!\!\!\!\!\sum_{n=-\infty}^\infty \!\sum_{\sigma=\pm
1}\![
c_{n,\sigma}\!(p_z,\!p_{_\perp}\!)\nonumber\\&&\!\!\!\!\times{\rm
e}\!^{-{\rm i}\varepsilon\!_{n,\sigma\!,1}
(p_z,p_{_\perp}\!)t}U_{n,\sigma}(p_z,p_{_\perp};\mbox{\boldmath$x$})+d^\dag_{n,\sigma}\!(p_z,p_{_\perp}\!)\nonumber\\
&&\!\!\!\!\times{\rm e}\!^{-{\rm
i}\varepsilon\!_{-n,-\sigma\!,-1}(-p_z,p_{_\perp}\!)t}V_{n,\sigma}(p_z,p_{_\perp};\mbox{\boldmath$x$})].
\end{eqnarray}This picture is called the rotation-interaction
picture, and denoted by the subscript $(ri)$.

To be complete, we have to add the Coulomb energy (\ref{3}) to the
interaction Hamiltonian. In the rotation-interaction picture, it is
to add the right hand side of
\begin{eqnarray}\!\!\!&&\!\!\!\!\!{H_c}_{(ri)}=\!\frac{1}{2}\!\int\!\! {\rm d}^3x\!\!\int\!\!{\rm
d}^3x' \Psi_{(ri)}^\dag\!(\mbox{\boldmath
$x$})\Psi_{(ri)}(\mbox{\boldmath $x$})\nonumber\\\!\!\!&&\times
\frac{\alpha}{|\mbox{\boldmath $x$}-\mbox{\boldmath $x$}'|}
\Psi_{(ri)}^\dag(\mbox{\boldmath $x$}')\Psi_{(ri)}(\mbox{\boldmath
$x$}').\end{eqnarray} to the right hand side of (\ref{s}). This is
important to maintain the gauge and Lorentz invariance of the
theory. Of course, in applications, we may ignore this term when the
correction of the first order of $\sqrt{\alpha}$ is enough.

\section{Application to the laser electron scattering\label{A5}}
Consider the laser electron collision, in the course a photon other
than those in the laser is created. This is a laser Compton
scattering. Suppose a circularly polarized laser propagates along
the $z$-direction, as shown in eq.(\ref{16}). A  beam of electrons
moves along the opposite direction. After collision a photon of wave
vector $\mbox{\boldmath $k$}'$ is created, and an electron of
momentum $\mbox{\boldmath $p$}$ in the beam is scattered to a state
of momentum $\mbox{\boldmath $p$}'$. Since the laser here is in a
specified mode of amplitude $A_c$, the label
$[a_{i'c}(\mbox{\boldmath $k$}')]$ of the 'vacuum state' in eqs.
(\ref{r}) and (\ref{t}) will be simplified to $A_c$, or further to
$A$. The vacuum state will be denoted by $|A;0\rangle$ in the
following. Consider the transition from the state
$|n,\sigma,p_z,0\rangle=c^\dag_{n,\sigma}(p_z,0)|A;0\rangle$ to the
state
$|\mbox{\boldmath$k$}',\mbox{\boldmath$e$}'_i;n',\sigma',p'_z,p'_{_\perp}\rangle
={a'_i}^\dag(\mbox{\boldmath$k$}')c^\dag_{n'\sigma'}(p'_z,p'_{_\perp})|A;0\rangle$.
The interaction matrix element is
\begin{eqnarray}&&\!\!\!\!\langle p'_{_\perp},p'_z,\sigma',n';\mbox{\boldmath$e$}'_i,\mbox{\boldmath$k$}'
|H'_{(ri)}|n,\sigma,p_z,0\rangle \nonumber\\&&\!\!\!={\rm e}^{{\rm
i}[\varepsilon_{n'\!\!,\,\sigma'\!\!,1}(p'_z\!,\,p'_{_\perp})+k'\!-\varepsilon_{n\!,\sigma,1}(p_z\!,\,0)]t}
\!\!\int\!\!
U^\dag_{n',\sigma'}(p'_z,p'_{_\perp};\mbox{\boldmath$x$})\nonumber\\&&\times\frac{e\mbox{\boldmath
$\alpha$}\cdot\mbox{\boldmath $e$}'^*_{ir} {\rm e}^{{-\rm
i}\mbox{\boldmath $k$}'_r\!\cdot\,\mbox{\boldmath
$x$}}}{\sqrt{(2\pi)^32k'}}U_{n,\sigma}(p_z,0;\mbox{\boldmath$x$}){\rm
d}^3x.\label{v}\end{eqnarray}From eq. (\ref{u}) we see
\begin{eqnarray}&&U_{n,\sigma}(p_z,0;\mbox{\boldmath$x$})=\delta_{n,0}U_{0,\sigma}(p_z,0;\mbox{\boldmath$x$})
\nonumber\\&&=\!\!\frac{1}{2\pi}{\rm e}^{{\rm
i}(p_z+\frac{eA}{2}Rk)z}\{1+kR[\alpha_x\cos kz+\alpha_y\sin kz
\nonumber\\&&+{\rm i}(\Sigma_y\cos kz-\Sigma_x\sin
kz)]/2\}u_\sigma(0) ,\label{v3}\end{eqnarray} the subscript $c$ of
$k_c$ is also omitted.  What we have to calculate is
\begin{eqnarray}&&\Theta_i\!=\!\int_0^\infty\!\!\!\!\!\rho{\rm
d}\rho\int_0^{2\pi}\!\!\!\!\!{\rm d}\varphi\;
u^\dag_{\sigma'}(0)\{{\rm i}^{-n'}{\rm J}_{n'}(p'_{_\perp}\rho'){\rm
e}^{{\rm i}n'\varphi'}{\rm P}'_+\nonumber\\&&+{\rm
i}^{-(n'-\sigma')}{\rm J}_{n'-\sigma'}(p'_{_\perp}\rho'){\rm
e}^{{\rm i}(n'-\sigma')\varphi'}{\rm
P}'_-\}\Xi_iu_\sigma(0)\nonumber\\&&\times\!\!\!\!\!\!\sum_{n"=-\infty}^\infty\!\!\!{\rm
i}^{-n"}{\rm e}^{{\rm i}n"\varphi\!_{_{k'_r}}}{\rm
J}_{n"}(k'_{_\perp}\rho){\rm e}^{-{\rm
i}n"\varphi},\end{eqnarray}with ${\rm
P}'_\pm\equiv\frac{1\pm\sigma'\Sigma_z}{2}$, and\begin{widetext}
\begin{eqnarray}\Xi_i&\equiv&\left\{1+\frac{kR'}{2}\left[\alpha_x\cos kz+\alpha_y\sin kz-
{\rm i}(\Sigma_y\cos kz-\Sigma_x\sin
kz)\right]\right\}\alpha_i\nonumber\\&\times&\left\{1+\frac{kR}{2}\left[\alpha_x\cos
kz+\alpha_y\sin kz+{\rm i}(\Sigma_y\cos kz-\Sigma_x\sin
kz)\right]\right\},\end{eqnarray}\end{widetext}$i=x,y,z.$
$\varphi\!_{_{k'_r}}=\varphi\!_{_{k'}}+kt$ is the longitude of
$\mbox{\boldmath$k$}'_r$, it is the angle between the projection of
$\mbox{\boldmath$k$}'_r$ on the $x-y$ plane and the $x-$axis.
$\varphi\!_{_{k'}}$ is the longitude of $\mbox{\boldmath$k$}'$,
which is time independent.  The calculation is elementary but
tedious. The result is\begin{widetext}
\begin{eqnarray}\Theta_i=\frac{2\pi}{p'_{_\perp}}\delta(p'_{_\perp}-k'_{_\perp})
{\rm e}^{{\rm
i}p'_{_\perp}R'\sin(kz-\varphi\!_{_{k'_r}})}\sqrt{\frac{1}{4EE'(E+m)(E'+m)}}(-1)^{n'}{\rm
e}^{{\rm i}n'\varphi\!_{_{k'_r}}}W_i,\end{eqnarray}
\begin{eqnarray}W_x\!\!\!\!&=&\!\!\!\!\left[\frac{k}{2}(p_z-E-m)(p'_z-E'-m)(R{\rm e}^{{\rm i}\sigma kz}+
R'{\rm e}^{-{\rm i}\sigma kz})-{\rm e}^{-{\rm
i}\sigma\varphi\!_{_{k'_r}}}p'_{_\perp}(E+m)\right]\delta_{\sigma,\sigma'}\nonumber\\
&+&\!\!\!\!\sigma\!\!\left[p_z(E'+m)-p'_z(E+m)-{\rm e}^{{\rm
i}\sigma\varphi\!_{_{k'_r}}}\frac{k}{2}(p_z-E-m)p'_{_\perp}(R{\rm
e}^{{\rm i}\sigma kz}+R'{\rm e}^{-{\rm i}\sigma
kz})\right]\!\!\delta_{\sigma,-\sigma'},\\W_y\!\!\!&=&\!\!\! {\rm
i}\left\{\sigma\left[\frac{k}{2}(p_z-E-m)(p'_z-E'-m)(R'{\rm
e}^{-{\rm i}\sigma kz}-R{\rm e}^{{\rm i}\sigma kz})-{\rm e}^{-{\rm
i}\sigma\varphi\!_{_{k'_r}}}p'_{_\perp}(E+m)\right]\delta_{\sigma,\sigma'}\right.\nonumber\\
&+&\!\!\!\!\!\!\left.\left[p_z(E'+m)-p'_z(E+m)-{\rm e}^{{\rm
i}\sigma\varphi\!_{_{k'_r}}}\frac{k}{2}(p_z-E-m)p'_{_\perp}(R'{\rm
e}^{-{\rm i}\sigma kz}-R{\rm e}^{{\rm i}\sigma
kz})\right]\!\!\delta_{\sigma,-\sigma'}\right\},\end{eqnarray}\begin{eqnarray}W_z\!\!\!\!&=&\!\!\!\!
\left\{(E'+m)p_z+(E+m)p'_z+\frac{k^2RR'}{2}(p_z-E-m)(p'_z-E'-m)\right.
\nonumber\\
&-&\!\!\!\!\left.\frac{kp'_{_\perp}}{2}{\rm e}^{{\rm
i}\sigma(kz-\varphi\!_{_{k'_r}})}\!\!\left[(R\!\!+\!\!R')(E+m)\!\!-\!\!(R\!\!-\!\!R')p_z\right]\right\}\!\delta_{\sigma,\sigma'}
\!\!-\!\!
\left\{\left[\frac{k^2RR'}{2}(p_z\!\!-\!\!E\!\!-\!m)\!\!+\!\!E+m\right]\!\!p_{_\perp}\!{\rm
e}^{{\rm
i}\sigma\varphi\!_{_{k'_r}}}\right.\nonumber\\\!\!\!\!&-&\!\!\!\!\left.\frac{k{\rm
e}^{{\rm i}\sigma
kz}}{2}\!\!\left[R(p_z-E-m)(p'_z+E'+m)\!\!-\!\!R'(p_z+E+m)(p'_z-E'-m)\right]\right\}\sigma\delta_{\sigma,-\sigma'}.
\end{eqnarray} \end{widetext}Take $\mbox{\boldmath$e$}'_1=\cos\theta(\cos\varphi\!_{_{k'}}\mbox{\boldmath$x$}_0
+\sin\varphi\!_{_{k'}}\mbox{\boldmath$y$}_0)-\sin\theta\mbox{\boldmath$z$}_0$
and $\mbox{\boldmath$e$}'_2=
-\sin\varphi\!_{_{k'}}\mbox{\boldmath$x$}_0+\cos\varphi\!_{_{k'}}\mbox{\boldmath$y$}_0$
to be a pair of orthonormal polarization vectors orthogonal to the
wave vector $\mbox{\boldmath$k$}'$, in which $\theta$ is the angle
between $\mbox{\boldmath$k$}'$ and the $z$ axis. We have
\begin{eqnarray}&&\mbox{\boldmath$\Theta$}\cdot\mbox{\boldmath$e$}'^*_{ir}=
\frac{2\pi}{p'_{_\perp}}\delta(p'_{_\perp}-k'_{_\perp}){\rm e}^{{\rm
i}p'_{_\perp}R'\sin(kz-\varphi\!_{_{k'_r}})}\nonumber\\&&\times\sqrt{\frac{1}{4EE'(E+m)(E'+m)}}(-1)^{n'}{\rm
e}^{{\rm i}n'\varphi\!_{_{k'_r}}}\nonumber\\&&\times
\left\{\delta_{\sigma,\sigma'}\sum_{\nu=0,\pm1}F^{(\nu)}_i{\rm
e}^{{\rm
i}\nu[k(z-t)-\varphi\!_{_{k'}}]}\right.\nonumber\\&&+\left.\delta_{\sigma,-\sigma'}\sum_{\nu=0,\pm1}G^{(\nu)}_i{\rm
e}^{{\rm i}\nu[k(z-t)-\varphi\!_{_{k'}}]}{\rm e}^{{\rm
i}\sigma\varphi\!_{_{k'_r}}}\right\},\label{v1}
\end{eqnarray}$i=1,2,$ in which
\begin{eqnarray}\left.\begin{array}{l}F^{(0)}_1=-\cos\theta
p'_\perp(E+m)-\sin\theta[(E'+m)p_z\\+\!(E+m)p'_z\!\!
+\!\frac{1}{2}k^2RR'(p_z\!-\!E\!-\!m)(p'_z\!-\!E'\!-\!m)],\\F^{(\sigma)}_1=\frac{k}{2}\{\cos\theta
R (p_z-\!E-\!m)(p'_z-\!E'-m)\\+\!\sin\theta
p'_\perp[(R\!+\!R')(E+m)\!-\!(R\!-\!R')p_z\},\\
F^{(-\sigma)}_1=\frac{k}{2}\cos\theta
R'(p_z-E-m)(p'_z-E'-m),\\G^{(0)}_1=\sigma\{\cos\theta[p_z(E'+m)-p'_z(E+m)]
\\+\sin\theta p'_\perp[\frac{k^2RR'}{2}(p_z-E-m)+E+m]\},\\
G^{(\sigma)}_1=-\frac{\sigma k}{2}\{\cos\theta
Rp'_\perp(p_z-E-m)\\+\sin\theta[R(p_z-E-m)(p'_z+E'+m)\\-R'(p_z+E+m)(p'_z-E'-m)]\}
,\\
G^{(-\sigma)}_1=-\frac{\sigma k}{2}\cos\theta
R'p'_\perp(p_z-E-m),\end{array}\!\!\!\!\right\}\end{eqnarray}
\begin{eqnarray}\left.\begin{array}{l}F^{(0)}_2=-{\rm i}\sigma p'_\perp(E+m),
\\F^{(\sigma)}_2=-{\rm i}\sigma\frac{kR}{2}(p_z-E-m)(p'_z-E'-m),\\
F^{(-\sigma)}_2={\rm
i}\sigma\frac{kR'}{2}(p_z-E-m)(p'_z-E'-m),\\G^{(0)}_2={\rm i}[p_z(E'+m)-p'_z(E+m)],\\
G^{(\sigma)}_2={\rm i}\frac{kR}{2}p'_\perp(p_z-E-m)
,\\
G^{(-\sigma)}_2=-{\rm
i}\frac{kR'}{2}p'_\perp(p_z-E-m).\end{array}\;\;\;\;\;\;\right\}\end{eqnarray}
Substituting (\ref{v1}) into (\ref{v}), using (\ref{v2}) once again,
we obtain\begin{eqnarray}&&\langle
p'_{_\perp},p'_z,\sigma',n';\mbox{\boldmath$e$}'_i,\mbox{\boldmath$k$}'
|H'_{(ri)}|n,\sigma,p_z,0\rangle
=\frac{e\delta(p'_{_\perp}-k'_{_\perp})}{\sqrt{(2\pi)^32k'}
p'_{_\perp}}\nonumber\\&&\times\sqrt{\frac{1}{4EE'(E+m)(E'+m)}}\sum_{{\cal
N}=-\infty}^\infty\sum_{\nu=0,\pm1}{\rm J}_{{\cal N}-\nu}(p'_\perp
R') \nonumber\\&&\times\left[\delta_{\sigma,\sigma'}F^{(\nu)}_i{\rm
e}^{-{\rm i}{\cal
N}\varphi\!_{_{k'}}}+\delta_{\sigma,-\sigma'}G^{(\nu)}_i{\rm
e}^{{\rm i}(\sigma-{\cal
N})\varphi\!_{_{k'}}}\right]\nonumber\\&&\times\delta
[p'_z-p_z+\frac{eA}{2}(R'-R)k+k'_z-{\cal N}
k]\nonumber\\&&\times{\rm e}^{{\rm i}n'(\varphi\!_{_{k'}}+\pi)}{\rm
e}^{{\rm i} [E'-E+\frac{eA}{2}(R'-R)k+k'-{\cal N}
k]t}.\end{eqnarray}Notice, terms containing  the factor
$\frac{\sigma}{2}-n$ in $\varepsilon_{n}$ and the factor
$\frac{\sigma'}{2}-n'$ in $\varepsilon_{n'}$ disappear due to
cancelations in the calculation. The matrix element is decomposed
into a series of terms with different time dependence, characterized
by an integer parameter ${\cal N}$. The resonance condition
\begin{equation}E'-E+\frac{eA}{2}(R'-R)k+k'-{\cal N}
k=0\label{w}\end{equation}is put on variables $E',E,p'\!_z,k',k$ and
on the parameter  ${\cal N}$, but not on $\sigma',\sigma,n'$ and
$n$. Therefore, transitions with same $E',E,p'\!_z,k',k$ and ${\cal
N}$, but different $\sigma',\sigma,n'$ and $n$, may have nonzero
probabilities simultaneously and be coherent with each other. This
point is important in the following. In the limit $A\rightarrow 0$,
this equation may be interpreted as energy conservation in the
collision, ${\cal N}$ is interpreted to be the number of photons in
laser absorbed during the collision. This interpretation is
questionable since its ignorance of coherence and interactions
between particle states. It becomes obvious when $A$ is finite.

In quantum field theory, people assumes that the interaction
switches on and off infinitely slow. This is the adiabatic
assumption. We take this assumption for the case with a laser
interaction, so that the Gell-Mann Low theorem\cite{gl} is
applicable to the equation (\ref{v4}). Eqs. (\ref{v5}) and (\ref{k})
show that at the limit $A\rightarrow 0$,
\begin{eqnarray}&&\!\!\!\!\!\!\!U_{n,\sigma}(p_z,p_{_\perp};\mbox{\boldmath$x$})\rightarrow
U^{(0)}_{n,\sigma}(p_z,p_{_\perp};\mbox{\boldmath$x$})\equiv\frac{{\rm
e}^{{\rm
i}p_zz}}{2\pi}\nonumber\\&&\!\!\!\!\!\!\!\!\!\times\!\!\left[{\rm
i}^n{\rm J}_n(p_{_\perp}\rho){\rm e}^{-{\rm i}n\varphi}{\rm
P}_+\!\!+\!{\rm i}^{n-\sigma}{\rm J}_{n-\sigma}(p_{_\perp}\rho){\rm
e}^{-{\rm i}(n-\sigma)\varphi}{\rm
P}_-\!\!\right]\!\!u_\sigma(0).\nonumber\\\end{eqnarray} This is a
solution of Dirac equation for a free electron, and
\begin{eqnarray}\sum_{n=-\infty}^\infty\frac{{\rm e}^{{\rm i}n\varphi\!_p}}
{\sqrt{2\pi}}U^{(0)}_{n,\sigma}(p_z,p_{_\perp};\mbox{\boldmath$x$})=\frac{1}{\sqrt{(2\pi)^3}}{\rm
e}^{{\rm i}\mbox{\boldmath$p$}\cdot\,\mbox{\boldmath$x$}}\;u
\end{eqnarray}is a plane wave solution of Dirac equation for a free
electron of momentum
$\mbox{\boldmath$p$}=p_{_\perp}(\cos\varphi_p\mbox{\boldmath$x$}_0
+\sin\varphi_p\mbox{\boldmath$y$}_0)+p_z\mbox{\boldmath$z$}_0$, $u$
is defined by eq.(\ref{v6}). In the case of $p_{_\perp}=0$, we
have\begin{equation}U^{(0)}_{n,\sigma}(p_z,0;\mbox{\boldmath$x$})=\frac{\delta_{n,0}}{2\pi}{\rm
e}^{{\rm i}p_zz}u.\label{w1}\end{equation} It is nonzero only when
$n=0$. In this case, it is already a plane wave of electron with
momentum $\mbox{\boldmath$p$}=p_z\mbox{\boldmath$z$}_0$.

Suppose a free electron of momentum
$\mbox{\boldmath$p$}=p_z\mbox{\boldmath$z$}_0$ and spin $\sigma$
comes from remote past and meets a laser on the way. It evolves
according to Gell-Mann Low theorem into the state
$U_{n,\sigma}(p_z,0;\mbox{\boldmath$x$})$. The above analysis shows,
this state may transit to a superposition $\sum_{n'=-\infty}^\infty
\frac{{\rm e}^{{\rm i}n'\varphi\!_{_p}}} {\sqrt{2\pi}}
U_{n',\sigma'}(p'_z,p'_{_\perp};\mbox{\boldmath$x$})$ of electron
states in the laser due to the electromagnetic interaction
$H'_{(ri)}$, and emit a photon of momentum $\mbox{\boldmath$k$}'$.
This superposition of electron states evolves once again in the
laser, into the state $\frac{1}{\sqrt{(2\pi)^3}}{\rm e}^{{\rm
i}\mbox{\boldmath$p$}'\cdot\,\mbox{\boldmath$x$}}\;u_{\sigma'}$ of a
free electron when goes to the remote future. In this process, the
initial state is
$|N,\mbox{\boldmath$k$}',\mbox{\boldmath$e$}'_i;\mbox{\boldmath$p$},\sigma\rangle
=\frac{1}{\sqrt{N!}}{a'_i}^{\dag
N}(\mbox{\boldmath$k'$})c^\dag_{0,\sigma}(p_z,0)|A;0\rangle$, with
$\mbox{\boldmath$p$}=p_z\mbox{\boldmath$z$}_0$. $N$ is the number of
photons of momentum $\mbox{\boldmath$k'$}\neq\mbox{\boldmath$k$}$
before the emission. The final state is
\begin{eqnarray}&&|N+1,\mbox{\boldmath$k$}',\mbox{\boldmath$e$}'_i;\mbox{\boldmath$p$}',\sigma'\rangle
=\frac{1}{\sqrt{(N+1)!}} \nonumber\\&&\times{a'_i}^{\dag
N+1}(\mbox{\boldmath$k'$})\!\!\!\!\sum_{n'=-\infty}^\infty\!\!\!\!
\frac{{\rm e}^{{\rm i}n'\varphi\!_{p'}}}
{\sqrt{2\pi}}c^\dag_{n',\sigma'}(p'_z,p'_{_\perp})|A;0\rangle
.\end{eqnarray}The interaction matrix element
is\begin{eqnarray}&&\langle
\sigma',\mbox{\boldmath$p$}';\mbox{\boldmath$e$}'_i,\mbox{\boldmath$k'$},N+1
|H'_{(ri)}|N,\mbox{\boldmath$k$}',\mbox{\boldmath$e$}'_i;\mbox{\boldmath$p$},\sigma\rangle
=\frac{e}{2\pi\sqrt{2k'}
}\nonumber\\&&\times\sqrt{\frac{N+1}{4EE'(E+m)(E'+m)}}\sum_{{\cal
N}=-\infty}^\infty\sum_{\nu=0,\pm1}{\rm J}_{{\cal N}-\nu}(p'_\perp
R') \nonumber\\&&\times\left[\delta_{\sigma,\sigma'}F^{(\nu)}_i{\rm
e}^{-{\rm i}{\cal
N}\varphi\!_{_{k'}}}+\delta_{\sigma,-\sigma'}G^{(\nu)}_i{\rm
e}^{{\rm i}(\sigma-{\cal
N})\varphi\!_{_{k'}}}\right]\nonumber\\&&\times\delta
[\mbox{\boldmath$p'+k'-p$}-{\cal N}
\mbox{\boldmath$k$}+\frac{eA}{2}(R'-R)\mbox{\boldmath$k$}]\nonumber\\&&\times{\rm
e}^{{\rm i} [E'-E+\frac{eA}{2}(R'-R)k+k'-{\cal N}
k]t},\end{eqnarray} in which\begin{eqnarray}&&\delta
[\mbox{\boldmath$p'+k'-p$}-{\cal N}
\mbox{\boldmath$k$}+\frac{eA}{2}(R'-R)\mbox{\boldmath$k$}]\nonumber\\&&=\frac{\delta(p'_{_\perp}-k'_{_\perp})}{p'_{_\perp}}
\delta(\varphi\!_{_{k'}}+\pi-\varphi\!_{_{p'}})\nonumber\\&&\times\delta
[p'_z-p_z+\frac{eA}{2}(R'-R)k+k'_z-{\cal N} k]
\end{eqnarray}is a 3-dimensional $\delta$-function.

In the first order perturbation, the transition amplitude of the
process is\begin{widetext}\begin{eqnarray}&&\langle
\sigma',\mbox{\boldmath$p$}';\mbox{\boldmath$e$}'_i,\mbox{\boldmath$k'$},N\!\!+\!\!1
|T|N,\mbox{\boldmath$k$}',\mbox{\boldmath$e$}'_i;\mbox{\boldmath$p$},\sigma\rangle
\!=\!\!-{\rm i}\!\!\int_{-\infty}^\infty\!\!\!\!\langle
\sigma',\mbox{\boldmath$p$}';\mbox{\boldmath$e$}'_i,\mbox{\boldmath$k'$},N\!\!+\!\!1
|H'_{(ri)}|N,\mbox{\boldmath$k$}',\mbox{\boldmath$e$}'_i;\mbox{\boldmath$p$},\sigma\rangle
{\rm d}t\!=\!\!-{\rm i}\frac{e}{\sqrt{2k'} }\nonumber\\&&
\times\sqrt{\frac{N+1}{4EE'(E+m)(E'+m)}}\sum_{{\cal
N}=-\infty}^\infty\!\sum_{\nu=0,\pm1}\!\!{\rm J}_{{\cal
N}-\nu}(p'_\perp R') \left[\delta_{\sigma,\sigma'}F^{(\nu)}_i{\rm
e}^{-{\rm i}{\cal
N}\varphi\!_{_{k'}}}+\delta_{\sigma,-\sigma'}G^{(\nu)}_i{\rm
e}^{{\rm i}(\sigma-{\cal
N})\varphi\!_{_{k'}}}\!\!\right]\nonumber\\&&\times\delta
[\mbox{\boldmath$p'+k'-p$}-{\cal N}
\mbox{\boldmath$k$}+\frac{eA}{2}(R'-R)\mbox{\boldmath$k$}]\delta[E'+k'-E-{\cal
N}
k+\frac{eA}{2}(R'-R)k].\end{eqnarray}\end{widetext}$\delta-$functions
give selection rules for non-zero transition probability. Besides
eq. (\ref{w}), we have
\begin{equation}\mbox{\boldmath$p'+k'-p$}-{\cal N}
\mbox{\boldmath$k$}+\frac{eA}{2}(R'-R)\mbox{\boldmath$k$}=0.\label{w5}\end{equation}In
the limit $A=0$, it means the usual momentum conservation of the
process. The $A-$dependent term shows the coherence effect of the
laser. Using these two selection rules we obtain
\begin{eqnarray}k'\!\!=\!\!\frac{{\cal N} k(E-p_z)}{E+{\cal N} k+\frac{eA}{2}Rk-(p_z
+{\cal N} k+\frac{eA}{2}Rk)\cos\theta}\, .\end{eqnarray}For given
incident electrons and laser, this formula gives the direction
dependence of the energy of emitted photon.  For ${\cal N}=1$ and in
the limit $A=0$, it reduces to the Compton formula for the usual
Compton scattering\cite{c1,c2}.

According to the collision theory, the transition probability
is\cite{we,l}
\begin{eqnarray}\!\!\!P\!=\!|\langle
\sigma',\mbox{\boldmath$p$}';\mbox{\boldmath$e$}'_i,\mbox{\boldmath$k'$},N\!\!+\!\!1
|T|N,\mbox{\boldmath$k$}',\mbox{\boldmath$e$}'_i;\mbox{\boldmath$p$},\sigma\rangle|^2{\rm
d}^3p'{\rm d}^3k'.\end{eqnarray}The easiest way to understand the
multiplication of ${\rm d}^3p'{\rm d}^3k'$ is by the method of box
normalization. The squares of $\delta-$functions are handled
by\begin{eqnarray}&&\delta[\Phi(\zeta)]{\rm
d}\zeta=\delta[\Phi(\zeta)]{\rm d}\Phi\frac{{\rm d}\zeta}{{\rm
d}\Phi}=\frac{{\rm d}\zeta}{{\rm d}\Phi},\nonumber\\&&\mbox{under
the condition of
$\Phi(\zeta)=0$},\end{eqnarray}and\begin{eqnarray}\delta(0)=\frac{1}{2\pi}\int_{-\infty}^\infty{\rm
d}\xi=\frac{\Delta\xi}{2\pi},\;\;\;\;\;\mbox{for macroscopic $\Delta
\xi$.}\end{eqnarray} The transition probability per-unit time in
unit volume and unit solid angle of $\mbox{\boldmath$k$}'$
is\begin{eqnarray}&&\!\!\!\!\!\!\frac{\partial^5
P}{\partial^3x\partial t\partial\Omega_{k'}}=\frac{\alpha
k'}{(2\pi)^3}\frac{N+1}{4EE'(E+m)(E'+m)}\nonumber\\&&\!\!\!\!\!\!\times\frac{E'k'}{{\cal
N} k(E-p_z)} \!\!\sum_{{\cal N}=-\infty}^\infty\!\!
\left|\sum_{\nu=0,\pm1}\!\!\!\!{\rm J}_{{\cal N}-\nu}(p'_\perp
R')\!\! \left[\delta_{\sigma,\sigma'}F^{(\nu)}_i{\rm e}^{-{\rm
i}{\cal
N}\varphi\!_{_{k'}}}\right.\right.\nonumber\\&&\!\!\!\!\!\!\left.\left.+\delta_{\sigma,-\sigma'}G^{(\nu)}_i{\rm
e}^{{\rm i}(\sigma-{\cal
N})\varphi\!_{_{k'}}}\!\!\right]\right|^2.\end{eqnarray}

In the problem of collision between particles, the cross section of
the target particle is measured and calculated. It is the
probability of occurrence of a given kind collision when there is an
incident particle comes across a unit area. Our problem is the
collision between an electron and a laser. The corresponding
quantity to be measured and calculated is the probability of
creating a photon in a volume of the laser by an electron coming
across a unit area. We call it the cross section of that volume of
the laser. In the remote past, the incident electron was in the
state (\ref{w1}). There is a piece of probability of amount
$|p_z|/[E(2\pi)^2]$ passing across a unit area in the $x-y$ plane
per unit time. In our unit system, the compton wavelength of an
electron is $m^{-1}$. In the unit of $m^{-2}$, the differential
cross section of a piece of laser of volume $m^{-3}$, in which a
photon of momentum $\mbox{\boldmath$k$}'$ is emitted in a unit solid
angle by an incident electron, is
\begin{eqnarray} &&\!\!\!\!\!\!\frac{{\rm d} \Sigma}{{\rm
d}\Omega_{k'}}=\frac{\alpha k'^2(N+1)}{8\pi m {\cal N} k
|p_z|(E-p_z)(E+m)(E'+m)}
\nonumber\\&&\!\!\!\!\!\!\times\!\!\!\!\sum_{{\cal
N}=-\infty}^\infty\! \left|\sum_{\nu=0,\pm1}\!\!\!\!{\rm J}_{{\cal
N}-\nu}(p'_\perp R')\!\! \left[\delta_{\sigma,\sigma'}F^{(\nu)}_i\!+
\delta_{\sigma,-\sigma'}G^{(\nu)}_i{\rm e}^{{\rm
i}\sigma\varphi\!_{_{k'}}}\!\!\right]\!\right|^2.\nonumber\\\label{w2}\end{eqnarray}

The cross section is decomposed into an incoherent sum of partial
cross sections, each associates an integer ${\cal N}$, which may be
imagined to be the number of photons in the laser participating the
collision. A partial cross section is proportional to an absolute
square of a superposition of three partial waves each contains a
Bessel function factor. It forms a characteristic diffraction
pattern depending on the integer ${\cal N}$, to be verified by
experiments. This result seems to be an interpretation of the
multi-photon process seen in experiments\cite{v,a}. However, the
physics shown in our derivation is not a multi-photon process, but
is an effect originated by the electron wave distortion in a laser.
It is exactly handled for the case of circularly polarized laser. In
the collision, only one photon is emitted directly by an electron.
It is handled by perturbation. This is justified by the smallness of
the fine structure constant $\alpha$. The precision of the
derivation here is therefore as high as those in the usual quantum
electrodynamics in vacuum.

\begin{figure}\includegraphics[width=8cm]{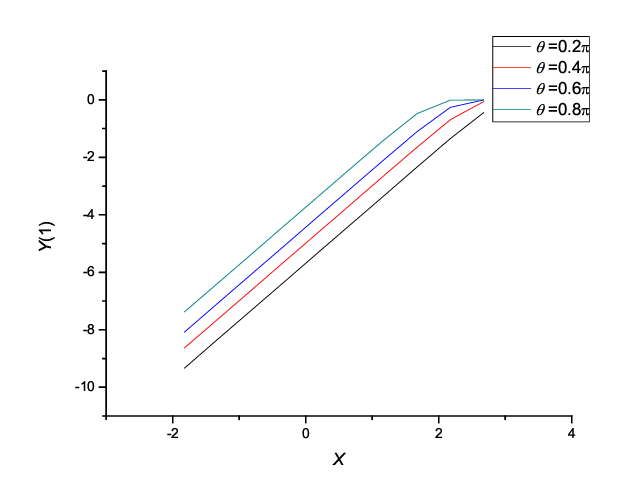}
\caption{Relations between $Y(1)$ and $X$, in which $E/m=3\times
10^2$ and $k/m=3.09\times 10^{-6}$, $\theta$ is the angle between
the wave vector $\mbox{\boldmath$k$}'$ of the outgoing photon state
and the wave vector $\mbox{\boldmath$k$}$ of the incident
laser.}\label{fig1}
\end{figure}

\begin{figure}\includegraphics[width=8cm]{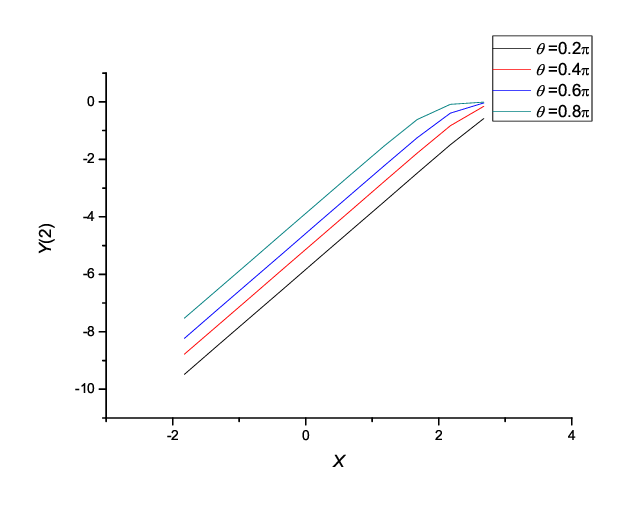}
\caption{Relations between $Y(2)$ and $X$, in which $E/m=3\times
10^2$ and $k/m=3.09\times 10^{-6}$, $\theta$ is the angle between
the wave vector $\mbox{\boldmath$k$}'$ of the outgoing photon state
and the wave vector $\mbox{\boldmath$k$}$ of the incident
laser.}\label{fig2}
\end{figure}

To compare this result with the Klein-Nishina formula\cite{kn,n} for
the usual Compton scattering is interesting. Take $N=0$ and ${\cal
N}=1$, calculate the average cross section
\begin{equation}\overline{\frac{{\rm d}\Sigma}{{\rm
d}\Omega_{k'}}}(i)=\frac{1}{2}\sum_{\sigma=\pm1}\sum_{\sigma'=\pm1}\frac{{\rm
d} \Sigma}{{\rm d}\Omega_{k'}}\label{w3}\end{equation} from the
formula (\ref{w2}). The result is still dependent on the
polarization $i$ of the outgoing photon state. On the other hand,
transform the Klein-Nishina formula into a coordinate frame, in
which an electron moves and makes a head-on collision with a target
photon. It may be easily done by Lorentz and gauge
transformations\cite{we,m}. In a laser of amplitude $A$ and
frequency $k$, the number of photons in a volume $m^{-3}$ is
$\frac{k}{m}(\frac{eA}{m})^2/(4\pi\alpha)$. Multiplying the
transformed Klein-Nishina formula by this number, we obtain the
total average cross section
\begin{eqnarray}&&\left.\overline{\frac{{\rm d}\Sigma}{{\rm
d}\Omega_{k'}}}(i) \,\right|_{KN}=\frac{\alpha(eAk')^2}{16\pi m
k|p_z|(E-p_z)}\left[\frac{(E-p_z)k}{(E-p_z\cos\theta)k'}\right.\nonumber\\&&\left.+\frac{(E-p_z\cos\theta)k'}{(E-p_z)k}
-2+4|\mbox{\boldmath$e$}\cdot\mbox{\boldmath$e$}'_i|^2\right]\label{w4}\end{eqnarray}of
an aggregate of non-coherent photons in a volume $m^{-3}$ for a
head-on incident electron. The unit of the cross section here is
again $m^{-2}$. Numerical calculation shows, for given $E,k,i$ and
$A$, the output from (\ref{w3}) with ${\cal N}=1$ is smaller than
that from (\ref{w4}). The coherence of photon states in the laser
depresses the transition probability. This depression in turn offers
a way for measuring the coherent amplitude $A$ of a photon beam. The
difference of outputs approaches zero rapidly when $A\rightarrow 0$.
It is reasonable, and may be regarded as a check of our derivation
here. Examples for $E/m=3\times 10^2$ and $k/m=3.09\times 10^{-6}$
are shown in  fig.\ref{fig1}  and fig.\ref{fig2},  relations between
\begin{eqnarray}&&\!\!\!\!\!\!\!\!Y(i)\equiv\log_{10}\left[\left(\left.\overline{\frac{{\rm d}\Sigma}{{\rm
d}\Omega_{k'}}}(i) \,\right|_{KN}-\overline{\frac{{\rm
d}\Sigma}{{\rm d}\Omega_{k'}}}(i)\right)/\left.\overline{\frac{{\rm
d}\Sigma}{{\rm d}\Omega_{k'}}}(i)
\,\right|_{KN}\right]\nonumber\\&&\!\!\!\!
\!\!\!\!\mbox{and}\;\;\;\; X\equiv\log_{10}(eA/m)\;\;\;\;\mbox{are
plotted.}\nonumber\end{eqnarray}

For an arbitrarily given polarization
$\mbox{\boldmath$e$}'=\sum_{i=1,2}c_i\mbox{\boldmath$e$}'_i$ of the
created photon, with $|c_1|^2+|c_2|^2=1$, the cross section
(\ref{w2}) becomes
\begin{eqnarray} &&\!\!\!\!\frac{{\rm d} \Sigma}{{\rm
d}\Omega_{k'}}=\frac{\alpha k'^2(N+1)}{8\pi m {\cal N} k
|p_z|(E-p_z)(E+m)(E'+m)}
\nonumber\\&&\!\!\!\!\times\!\!\!\sum_{{\cal
N}=-\infty}^\infty\left|
\left[\delta_{\sigma,\sigma'}\mbox{\boldmath$e$}'^*\cdot\mbox{\boldmath${\cal
F}$}\!\!\!_{_{\cal N}}
+\delta_{\sigma,-\sigma'}\mbox{\boldmath$e$}'^*\cdot\mbox{\boldmath${\cal
G}$}\!_{_{\cal N}}\right]\right|^2,\label{wa}\end{eqnarray} in which
\begin{eqnarray}\left.\begin{array}{c}\mbox{\boldmath${\cal
F}$}\!\!_{_{\cal N}}=\sum_{\nu =0,\pm
1}\mbox{\boldmath$F$}^{(\nu)}{\rm J}_{{\cal N}-\nu}(p'_\perp R')
\\\mbox{\boldmath${\cal G}$}\!_{_{\cal N}}=\sum_{\nu =0,\pm
1}\mbox{\boldmath$G$}^{(\nu)}{\rm J}_{{\cal N}-\nu}(p'_\perp
R'),\end{array}\right\}
\end{eqnarray}with
\begin{eqnarray}\mbox{\boldmath$F$}^{(\nu)}\!\!\equiv\!\!
\sum_{i=1,2}\!F_i^{(\nu)}\mbox{\boldmath$e$}'_i,\;\;\;\;
\mbox{\boldmath$G$}^{(\nu)}\!\!\equiv\!\!\sum_{i=1,2}\!G_i^{(\nu)}{\rm
e}^{{\rm
i}\sigma\varphi\!_{_{k'}}}\mbox{\boldmath$e$}'_i.\end{eqnarray} For
definite $\sigma$, $\sigma'$ and ${\cal N}$, the polarization of the
outgoing photon is also definite. It is
$\mbox{\boldmath$e$}'_f\equiv\mbox{\boldmath${\cal F}$}\!_{_{\cal
N}}/{\cal F}\!_{_{\cal N}}$ for $\sigma'=\sigma$ or
$\mbox{\boldmath$e$}'_g\equiv\mbox{\boldmath${\cal G}$}\!_{_{\cal
N}}/{\cal G}\!_{_{\cal N}}$ for $\sigma'=-\sigma$, with ${\cal
F}\!_{_{\cal N}}\equiv\sqrt{\mbox{\boldmath${\cal F}$}^*_{_{\cal
N}}\cdot\mbox{\boldmath${\cal F}$}\!_{_{\cal N}}}$ and ${\cal
G}\!_{_{\cal N}}\equiv\sqrt{\mbox{\boldmath${\cal G}$}^*_{_{\cal
N}}\cdot\mbox{\boldmath${\cal G}$}\!_{_{\cal N}}}$. For an
arbitrarily given polarization $\mbox{\boldmath$e$}'$ of the
outgoing photon, the average cross section (\ref{w3}) becomes
\begin{eqnarray}\overline{\frac{{\rm d}\Sigma}{{\rm
d}\Omega_{k'}}}(\mbox{\boldmath$e$}')&=&\frac{\alpha
k'^2(N+1)}{16\pi m {\cal N} k
|p_z|(E-p_z)(E+m)(E'+m)}\nonumber\\&\times&\sum_{{\cal
N}=-\infty}^\infty \sum_{\sigma=\pm
1}\mbox{\boldmath$e$}'^*\cdot\mbox{\boldmath${\cal Z}$}_{_{\cal
N}}\cdot\mbox{\boldmath$e$'} ,\end{eqnarray}in which
\begin{equation}\mbox{\boldmath${\cal
Z}$}_{_{\cal N}}\equiv\mbox{\boldmath${\cal F}$}_{_{\cal
N}}\mbox{\boldmath${\cal F}$}_{_{\cal N}}^* + \mbox{\boldmath${\cal
G}$}_{_{\cal N}}\mbox{\boldmath${\cal G}$}_{_{\cal
N}}^*\end{equation}is a tensor. If the polarization of the outgoing
photon is not measured, the average cross section is
\begin{eqnarray}&&\overline{\frac{{\rm d}\Sigma}{{\rm
d}\Omega_{k'}}}=\frac{\alpha k'^2(N+1)}{16\pi m {\cal N} k
|p_z|(E-p_z)(E+m)(E'+m)}\nonumber\\&&\times\sum_{{\cal
N}=-\infty}^\infty \sum_{\sigma=\pm 1}({\cal F}_{_{\cal N}}^2 +{\cal
G}_{_{\cal N}}^2).\end{eqnarray} Suppose the incident electron and
the outgoing photon are polarized, with certain quantum numbers
$\sigma$ and $\mbox{\boldmath$e$}'$. The state of outgoing electron
is a superposition
$a_1|\mbox{\boldmath$k$}',\mbox{\boldmath$e$}';n',\sigma_1',p'_z,p'_{_\perp}\rangle+a_2
|\mbox{\boldmath$k$}',$\mbox{\boldmath$e$}'$;n',\sigma'_2,p'_z,p'_{_\perp}\rangle$
of states with $\sigma'_1=\sigma$, $\sigma'_2=-\sigma$, and
$|a_1|^2+|a_2|^2=1$ in general. The cross section (\ref{wa}) becomes
\begin{eqnarray} &&\!\!\!\!\frac{{\rm d} \Sigma}{{\rm
d}\Omega_{k'}}=\frac{\alpha k'^2(N+1)}{8\pi m {\cal N} k
|p_z|(E-p_z)(E+m)(E'+m)}\nonumber\\&&\!\!\!\!\times\!\!\!
\sum_{{\cal
N}=-\infty}^\infty\mbox{\boldmath$e$}'^*\cdot\mbox{\boldmath${\cal
Z}$}_{_{\cal N}}\cdot\mbox{\boldmath$e$'}\left|
\chi'^\dag\chi_{_{\cal N}}\right|^2,\end{eqnarray}in which
\begin{eqnarray}\chi'\!\!=\!\!\left[\begin{array}{c}a_1\\a_2\end{array}\right]
\;\;\;\mbox{and}\;\;\;\chi_{_{\cal
N}}\!\!=\!\!\frac{1}{\sqrt{\mbox{\boldmath$e$}'^*\cdot\mbox{\boldmath${\cal
Z}$}_{_{\cal
N}}\cdot\mbox{\boldmath$e$'}}}\!\!\left[\begin{array}{c}\mbox{\boldmath$e$}'^*\cdot\mbox{\boldmath${\cal
F}$}\!_{_{\cal N}}\\\mbox{\boldmath$e$}'^*\cdot\mbox{\boldmath${\cal
G}$}\!_{_{\cal N}}\end{array}\right]\end{eqnarray}are normalized
spinors in the $[\chi_\sigma]$ representation. Since the cross
section is zero when $\chi'$ is orthogonal to $\chi_{_{\cal N}}$,
the outgoing electron is polarized. Its spin state is described by
the spinor $\chi_{_{\cal N}}$.
\section{The case of large ${\cal N}$\label{A6}}
The Bessel function in equation (\ref{w2}) may be easily worked out
by its series expression when the order and the argument are small.
On the other hand, from selection rules (\ref{w}) and (\ref{w5}) we
see $E'-p'_z=E-p_z-k'(1-\cos\theta)$. The argument of the Bessel
function here is therefore\begin{eqnarray}&&\!\!\!\!\!\!p'_\perp
R'=\sin\theta\frac{{\cal N}(E-p_z)}{E+({\cal
N}k+\frac{eA}{2}Rk)(1-\cos\theta)-p_z\cos\theta}\nonumber\\
&&\!\!\!\!\!\!\times\frac{eA}{E-p_z-k'(1-\cos\theta)}\nonumber\\
&&\!\!\!\!\!\!=\frac{{\cal N} eA
\sin\theta}{E+\frac{eA}{2}Rk(1-\cos\theta)-p_z\cos\theta}={\cal S
N}\, ,\end{eqnarray}${\cal S}$ is independent of  ${\cal N}$. When
${\cal N}$ becomes large, the order and the argument of the Bessel
function become large simultaneously. The asymptotic form for this
case is well known\cite{wa,wg}. For our case, $0<{\cal S}\leq1$, we
have ${\cal S}=\sec\!{\rm h} \,\xi,\, \xi$ is a real number. The two
term asymptotic form of the Bessel function is
\begin{eqnarray}&&{\rm J}_{{\cal N}\pm\nu}({\cal N}\sec\!{\rm h} \,\xi)
\simeq{\rm J}_{\cal N}({\cal N}\sec\!{\rm h}\,
\xi)\nonumber\\&&\sim\frac{{\rm e}^{{\cal N}(\tanh
\xi-\xi)}}{\sqrt{2{\cal
N}\pi\tanh\xi}}\left[1-\left(\frac{1}{8}-\frac{5}{24}\coth^2\xi\right)\frac{1}{{\cal
N}\tanh\xi}\right],\nonumber\\&& \mbox{for}\,\nu=0,\pm1
\,\mbox{and}\,{\cal N} \rightarrow\infty.\end{eqnarray}This
asymptotic form makes us able to consider the case of large ${\cal
N}$. When other parameters are fixed, ${\cal N}$ may be fixed by
maximizing the cross section.
\begin{table}\caption{Some numerical results of
average cross sections calculated by use of (\ref{w3}) with
$\theta=3.14$ and $eA/m=1.50\times10^{-2}$}\begin{tabular}{ccccc}
\hline  ${\cal N}$& $E/m$ &$k/m$&$k'/m$&$\overline{\frac{{\rm
d}\Sigma}{{\rm d}\Omega_{k'}}}(1)$\\\hline
1&$7.0\times10^3$&$3.09\times10^{-6}$&2.648&$2.27\times10^{-9}$\\\hline
2&$7.0\times10^3$&$3.09\times10^{-6}$&5.28&$9.99\times10^{-13}$\\\hline
3&$7.0\times10^3$&$3.09\times10^{-6}$&7.89&$4.18\times10^{-16}$\\\hline
\end{tabular}\label{tab1}\\
\caption{Some numerical results of average cross sections calculated
by use of (\ref{w3}) with $\theta=3.14$ and
$eA/m=10.5$}\begin{tabular}{ccccc} \hline ${\cal N}$& $E/m$
&$k/m$&$k'/m$&$\overline{\frac{{\rm d}\Sigma}{{\rm
d}\Omega_{k'}}}(1)$\\\hline
1&$7.0\times10^3$&$4.43\times10^{-9}$&$7.76\times10^{-5}$&$4.17\times10^{-8}$\\\hline
2&$7.0\times10^3$&$4.43\times10^{-9}$&$1.55\times10^{-4}$&$3.67\times10^{-9}$\\\hline
3&$7.0\times10^3$&$4.43\times10^{-9}$&$2.33\times10^{-4}$&$3.05\times10^{-10}$\\\hline
523&$7.0\times10^3$&$4.43\times10^{-9}$&1.90&$1.08\times10^{-8}$\\\hline\end{tabular}\label{tab2}
\end{table}

Some numerical results are listed in the tables, one case occupies
one row. For an example, the first row of table \ref{tab1} shows the
case 1 with ${\cal N}=1$. In this case, in the usual units, the
energy of the incident electron is
$7\times10^3\times0.51\mbox{MeV}=3.57$GeV, the energy of a photon in
the head-on laser is
$3.09\times10^{-6}\times0.51\mbox{MeV}=1.576\mbox{eV}$, the energy
of the emitted photon is $2.648\times0.51\mbox{MeV}=1.35\mbox{MeV}$,
and the average differential cross section for emitting a photon of
the polarization 1 per volume $\lambda_c^3$ of the laser is
$2.27\times10^{-9}(\lambda_c^2)$, in which
$\lambda_c=\hbar/mc=3.86\times10^{-13}$m is the Compton wave length
of the electron. The last row of table \ref{tab2} shows another
example. It is a case of large ${\cal N}=523$. The energy of the
incident electron is again 3.57GeV, the energy of a photon in the
laser is now $4.43\times10^{-9}\times0.51\mbox{MeV}=2.259$meV, the
energy of the emitted photon is
$1.9\times0.51\mbox{MeV}=0.97\mbox{MeV}$, and the average
differential cross section for emitting a photon of the polarization
1, per volume $\lambda_c^3$ of the laser, is
$1.08\times10^{-8}(\lambda_c^2)$. The angle $\theta$ between the
emitting direction and the direction of the laser propagation is
3.14 for all these cases. From the $k/m$ and $eA/m$ we may calculate
the energy flux of the laser. It is $10^{19}$W/m$^2$ for all these
seven cases, and is reachable by present techniques.

\section{The possibility of making a $\gamma$-ray laser\label{A7}}
In all cases shown in table \ref{tab1} and the case shown in the
last row of the table \ref{tab2} we see, energies of the emitted
photons are in the $\gamma-$ray range. The possibility of
constructing a $\gamma$-ray laser by the collision between high
energy electrons and an usual laser is worthy to consider. Suppose
the electron-laser collision happens in a tube of cross section $S$
and length $L$, with $\sqrt{S}\ll L$. Electrons come from an
accelerator, enter the tube at one of its ends, say end 1; then move
to another end, say end 2, and exit from the tube there. The laser
enters the tube at end 2, then propagates along an opposite
direction in the tube to end 1, and exits there. Denote the distance
between a point in the tube and the end 1 by $l$. The volume ${\rm
d}V$ of a space in the tube between two sections at $l$ and $l+{\rm
d}l$ respectively is $S{\rm d}l$. The total average differential
cross section of the laser in this space for an electron of energy
$E$ emitting a photon of polarization $i$ and wave vector
$\mbox{\boldmath $k$}'$ is $S\overline{\frac{{\rm d}\Sigma}{{\rm
d}\Omega_{k'}}}(i)\frac{{\rm d}l}{\lambda_c}$. The probability of
emitting a photon of this kind by an electron passing through the
space ${\rm d}V$ is $\overline{\frac{{\rm d}\Sigma}{{\rm
d}\Omega_{k'}}}(i)\frac{{\rm d}l}{\lambda_c}$. The number of photons
of this kind emitted by a burst of $n$ those electrons in the space
${\rm d}V$ is\begin{equation}{\rm d}N=[N(l)+1]a\frac{{\rm
d}l}{\lambda_c},\;\;\;\;\mbox{with}\;\;a=n\overline{\frac{{\rm
d}\Sigma}{{\rm
d}\Omega_{k'}}}(i)/[N(l)+1]\label{w6}\end{equation}During the
emission, the energy of the electron decreases. Suppose there are
some techniques being able to supply the electron energy (pump), and
maintain it to be the value $E$, the number $n$ of electrons having
energy $E$, and therefore the parameter $a$ in this equation, may be
a constant in the whole course. In this case, the solution of
equation (\ref{w6}) under the initial condition $N(0)=0$
is\begin{equation}N(l)={\rm e}^{al/\lambda_c}-1.\end{equation}For
$n\sim 1, \,\overline{\frac{{\rm d}\Sigma}{{\rm
d}\Omega_{k'}}}(i)/[N(l)+1]\sim 10^{-8},$ and $L\sim$ meters, $N(L)$
is almost infinity. This is the $\gamma-$ray amplification by the
stimulated emission of radiation.

The pump technique may be a way of accelerating electrons along the
tube. Since the change of electron energy distribution due to
emissions in the tube may be calculated by the present theory, the
way of acceleration for maintaining the original electron energy
distribution may be designed. Another way of pumping is to open more
entrances and exits on the wall along the tube, so that one may
substitute the bust of electrons with required energy distribution
for the old electron burst, which has been changed due to emissions.

The difficulty of finding transition schemes for amplifying
$\gamma-$rays is not serious. Many processes are suitable for this
purpose\cite{zg}. The difficulty of finding a resonator for
$\gamma-$ray amplification is truly serious. The tube suggested
above, if long enough, may play the role, which was played by the
resonator in the usual lasers. It makes us feel hopeful.
\section{Discussions\label{A8}}
Since the fine structure constant $\alpha$ is small, the
electron-photon interaction is weak, the quantum electrodynamics of
a system composed of a few electrons and photons in the vacuum may
be handled by perturbation. However, because of a huge number of
photons in the laser, the electron-laser interaction may be not
weak. The quantum electrodynamics could not be handled by
perturbation in general, if a laser is present. The way out of this
difficulty is to divide the interaction into two parts. One is the
interaction, by which an electron absorbs or emits a photon. This
part is still weak, and may be handled by perturbation. Another is
the interaction between an electron and a laser. This part may be so
strong, that the perturbation method breaks down. In this case, one
has to find other reliable methods to handle. We considered the
electron-laser collision, in which a photon other than those in the
laser is emitted. The electron laser interaction makes a distortion
of the electron wave, which is exactly solved for the circularly
polarized laser. The emission of photon by the electron is handled
by perturbation as usual. It is justified by the smallness of the
fine structure constant $\alpha$. The theory is as precise as that
for the usual quantum electrodynamics treatment of the
emission-absorption processes in vacuum. By use of this formulation
we see that there may be some phenomena indeed which were
interpreted to be the multi-photon processes\cite{v,a}. But
according to our analysis, they are phenomena originated by the
electron wave distortion, induced by the laser field. It is
therefore truly non-perturbative.

To realize this procedure, we divided the electromagnetic field into
two parts. One is the laser and another is the deviation of the
electromagnetic field from it, as did in equations (\ref{17}) and
(\ref{a}). The laser is a given classical field, and the deviation
from it is quantized. The quantization of electromagnetic field
around one of its classical solution is a generalization of its
usual quantization (around the vacuum), and is equivalent to it. In
the canonical quantization, to quantize a classical theory of
canonical variables $q_c$ and $p_c$, one substitutes q-numbers $q$
and $p$ for c-numbers $q_c$ and $p_c$, and applies the quantization
rule $qp-pq={\rm i}$ to them. Now for an arbitrary pair of c-number
functions $q_c(t)$ and $p_c(t)$ of time $t$, we may introduce the
variables $Q\equiv q-q_c(t)$ and $P\equiv p-p_c(t)$ instead of $q$
and $p$, to equivalently describe the system. They satisfy the same
quantization rule $QP-PQ={\rm i}$. The quantization based on new
pairs of variables is a generalization of the one based on old
pairs, and is equivalent to it. Though this substitution is almost
trivial, it does offer something new. Besides its application to the
electron laser collision here, we applied it to calculate the
residual interaction between nucleons on the basis of a relativistic
mean field solution for nuclear matter\cite{z2}.

The main skill in this work is to find a complete set of orthonormal
eigenfunctions for equation (\ref{l}). It is derived from Wolkow
solution\cite{w} of Dirac equation for an electron in a plane
electromagnetic wave. The result is used in the quantization of
electron field in our case. The same set also offers a general
solution of the relativistic quantum mechanics for an electron in a
circularly polarized electromagnetic field. It may be used in
researches of the electron motion in a laser as well. Among others,
it may give tips on how to shorten the electron bursts in the laser.
It is important in the techniques for laser acceleration of
electrons\cite{ch}.

Eq. (\ref{w2}) shows the existence of the stimulated emission in
electron laser collision. In section \ref{A7} we considered the
possibility of making a $\gamma-$ray laser by use of this process.
The result is exciting. It is possible to be realized with an
acceptable size.

\begin{center}{\bf ACKNOWLEDGMENTS}\end{center}

The work is supported by the Nature Science Foundation of China with
grant number 10875003.

\begin{center}{\bf APPENDIX: MAIN STEPS FOR DERIVING EQUATION (\ref{x})}\end{center}
They are:\begin{widetext}
\begin{eqnarray}\Xi&\equiv&\left\{1+\frac{k_cR}{2}\left[\alpha_x\cos k_cz+\alpha_y\sin k_cz-
{\rm i}(\Sigma_y\cos k_cz-\Sigma_x\sin
k_cz)\right]\right\}\nonumber\\&\times&\left\{1+\frac{k_cR'}{2}\left[\alpha_x\cos
k_cz+\alpha_y\sin k_cz+{\rm i}(\Sigma_y\cos k_cz-\Sigma_x\sin
k_cz)\right]\right\}\nonumber\\&=&1+\frac{k_c^2RR'}{2}(1-\alpha_z)+\frac{k_c}{2}(R+R')(\alpha_x\cos
k_cz+\alpha_y\sin k_cz)\nonumber\\&&+\frac{{\rm
i}k_c}{2}(R'-R)(\Sigma_y\cos k_cz-\Sigma_x\sin k_cz),\end{eqnarray}
\begin{eqnarray} \Theta\!\!&\equiv&\!\!\left\{{\rm i}^{-n}{\rm
J}_n(p_{_\perp}\rho'){\rm e}^{{\rm i}n\varphi'}{\rm P}_++{\rm
i}^{-(n-\sigma)}{\rm J}_{n-\sigma}(p_{_\perp}\rho'){\rm e}^{{\rm
i}(n-\sigma)\varphi'}{\rm
P}_-\right\}\Xi\nonumber\\&\times&\!\!\!\!\left\{{\rm i}^{n'}{\rm
J}_{n'}(p'_{_\perp}\rho''){\rm e}^{-{\rm i}n'\varphi''}{\rm P}'_+
+{\rm i}^{(n'-\sigma')}{\rm J}_{n'-\sigma'}(p'_{_\perp}\rho''){\rm
e}^{-{\rm
i}(n'-\sigma')\varphi''}{\rm P}'_- \right\}\nonumber\\
&=&\!\!\!\!\left[1+\frac{k_c^2RR'}{2}(1-\alpha_z)\right]\left\{\delta_{\sigma,\sigma'}{\rm
i}^{n'-n}\left[ {\rm J}_n(p_{_\perp}\rho'){\rm
J}_{n'}(p'_{_\perp}\rho''){\rm e}^{{\rm
i}(n\varphi'-n'\varphi'')}{\rm
P}_+\right.\right.\nonumber\\&+&\!\!\!\!\left.{\rm
J}_{n-\sigma}(p_{_\perp}\rho'){\rm
J}_{n'-\sigma}(p'_{_\perp}\rho''){\rm e}^{{\rm
i}(n-\sigma)\varphi'}{\rm e}^{-{\rm i}(n'-\sigma)\varphi''}{\rm
P}_-\right]+\delta_{\sigma,-\sigma'}{\rm
i}^{n'-n+\sigma}\nonumber\\&\times&\!\!\!\!\left.\left[{\rm
J}_n(p_{_\perp}\rho'){\rm J}_{n'+\sigma}(p'_{_\perp}\rho''){\rm
e}^{{\rm i}n\varphi'}{\rm e}^{-{\rm i}(n'+\sigma)\varphi''}{\rm
P}_++{\rm J}_{n-\sigma}(p_{_\perp}\rho'){\rm
J}_{n'}(p'_{_\perp}\rho''){\rm e}^{{\rm i}(n-\sigma)\varphi'}{\rm
e}^{-{\rm i}n'\varphi''}{\rm
P}_-\right]\right\}\nonumber\\&+&\!\!\!\!\frac{1}{2}\left[k_c(R+R')(\alpha_x\cos
k_cz+\alpha_y\sin k_cz)+{\rm i}k_c(R'-R)(\Sigma_y\cos
k_cz-\Sigma_x\sin k_cz)\right]\left\{\delta_{\sigma,\sigma'}{\rm
i}^{n'-n-\sigma}\right.\nonumber\\&\times&\!\!\!\!\left[{\rm
J}_n(p_{_\perp}\rho'){\rm J}_{n'-\sigma}(p'_{_\perp}\rho''){\rm
e}^{{\rm i}n\varphi'}{\rm e}^{-{\rm i}(n'-\sigma)\varphi''}{\rm
P}_--{\rm J}_{n-\sigma}(p_{_\perp}\rho'){\rm
J}_{n'}(p'_{_\perp}\rho''){\rm e}^{{\rm i}(n-\sigma)\varphi'}{\rm
e}^{-{\rm i}n'\varphi''}{\rm
P}_+\right]\nonumber\\&+&\!\!\!\!\delta_{\sigma,-\sigma'}{\rm
i}^{n'-n}\!\!\left[{\rm J}_n(p_{_\perp}\rho'){\rm
J}_{n'}(p'_{_\perp}\rho''){\rm e}^{{\rm
i}(n\varphi'-n'\varphi'')}{\rm
P}_-\right.\nonumber\\&-&\!\!\!\!\left.\left.{\rm
J}_{n-\sigma}(p_{_\perp}\rho'){\rm
J}_{n'+\sigma}(p'_{_\perp}\rho''){\rm e}^{{\rm
i}(n-\sigma)\varphi'}{\rm e}^{-{\rm i}(n'+\sigma)\varphi'')}{\rm
P}_+\right]\right\};
\end{eqnarray}and\begin{equation}\int_0^\infty\!\!\!\!
\int_0^{2\pi}\!\!\!\!u^\dag_\sigma(0)\Theta u_{\sigma'}(0)\rho{\rm
d}\rho{\rm d}\varphi
\!=\!\frac{\pi}{p_{_\perp}}\delta(p_{_\perp}-p_{_\perp}'){\rm
e}^{{\rm
i}[\frac{\sigma'-\sigma}{2}+n-n']k_cz}W,\end{equation}with\begin{eqnarray}
&&W\equiv u^\dag_\sigma(0)\{[2+
k_c^2RR'(1-\alpha_z)][\delta_{\sigma,\sigma'}{\rm
J}_{n-n'}(p_{_\perp}\Delta R)+\delta_{\sigma,-\sigma'}{\rm
J}_{n-n'-\sigma}(p_{_\perp}\Delta R)]\nonumber\\&&-k_c[(\Sigma
R\gamma_5+\Delta R)\Sigma_-\!+\!(\Sigma R\gamma_5-\Delta
R)\Sigma_+\!][\delta_{\sigma,\sigma'}({\rm
J}_{n-n'+\sigma}(p_{_\perp}\Delta R){\rm P}_-\!+\!{\rm
J}_{n-n'-\sigma}(p_{_\perp}\Delta R){\rm P}_+)\nonumber
\\&&+\delta_{\sigma,-\sigma'}({\rm
J}_{n-n'}(p_{_\perp}\Delta R){\rm P}_-+{\rm
J}_{n-n'-2\sigma}(p_{_\perp}\Delta R){\rm
P}_+)]\}u_{\sigma'}(0),\label{n}\end{eqnarray}\end{widetext}in which
$\Sigma R\equiv R+R',\Delta R\equiv
R'-R,\Sigma_\pm\equiv\frac{\Sigma_x\pm{\rm i}\Sigma_y}{2},
\gamma_5\equiv\gamma_1\gamma_2\gamma_3\gamma_4=-\alpha_i\Sigma_i,i=x,y,z$.
The formula (\ref{m}), formulae
\begin{eqnarray}\left.\begin{array}{l}\alpha_i\alpha_j+\alpha_j\alpha_i=2\delta_{i,j},\\
\Sigma_i\Sigma_j+\Sigma_j\Sigma_i=2\delta_{i,j},\end{array}\right\}\;\;\;\mbox{for}\;\;i,j=x,y,z,\end{eqnarray}
the orthogonal relations
\begin{eqnarray}\left.\begin{array}{l}\int_0^\infty{\rm J}_n(p_{_\perp}\rho)
{\rm J}_n(p'_{_\perp}\rho)\rho{\rm
d}\rho=\frac{1}{p_{_\perp}}\delta(p_{_\perp}-p'_{_\perp}),\\
\int_0^{2\pi}{\rm e}^{{\rm i}(\nu-\nu')\varphi}{\rm
d}\varphi=2\pi\delta_{\nu,\nu'},\end{array}\right\}
\end{eqnarray}and
the addition theorem
\begin{equation}\sum_{n'=-\infty}^\infty{\rm J}_{n+n'}(\xi_1)
{\rm J}_{n'}(\xi_2)={\rm J}_n(\xi_1-\xi_2)\end{equation}are used.
Now,\begin{equation}I=\frac{\pi}{p_{_\perp}}\delta(p_{_\perp}-p_{_\perp}')\int_{-\infty}^\infty{\rm
e}^{{\rm i}(\eta_ {n\!'}-\eta_n)z} W{\rm d}z,
\label{o}\end{equation}with
$\eta_n=p_z+\frac{e^2A_c^2}{2(E-p_z)}+(\frac{\sigma}{2}-n)k_c$ and
$\eta_{n'}=p'_z+\frac{e^2A_c^2}{2(E'-p'_z)}+(\frac{\sigma'}{2}-n')k_c$.
Since $W$ is $z$ independent, the nonzero condition for the integral
in eq. (\ref{o}) is $\eta_n= \eta_{n'}$. The nonzero conditions for
$I$ include $\varepsilon_n=\varepsilon_{n'},\eta_n= \eta_{n'}$ and
$p_{_\perp}=p'_{_\perp}$. After a careful analysis we see that they
mean $p_{_\perp}=p'_{_\perp}, p_z=p'_z,\tau=\tau'$, $\Delta R=0$,
$n=n'$ if $\sigma =\sigma'$, $n'=n-\sigma$ if $\sigma=-\sigma'$.
Since ${\rm J}_n(0)=\delta_{n,0}$ and
$u^\dag_\sigma(0)\alpha_zu_{\sigma
'}(0)=\frac{p_z}{E}\delta_{\sigma,\sigma'}$, we finally obtain
\begin{widetext}\begin{eqnarray}I&\!\!\!=&\!\!\!
\frac{4\pi^2}{p_\perp}\left[1+\frac{k^2R^2}{2}
\left(1-\frac{p_z}{E}\right)\right]\delta(p_{_\perp}-p'_{_\perp})
\delta(\eta-\eta')\delta_{n,n'}\delta_{\sigma,\sigma'}\delta_{\tau,\tau'}
\nonumber\\&\!\!\!=&\!\!\!\frac{4\pi^2}{p_\perp}\left[1+\frac{k^2R^2}{2}
\left(1-\frac{p_z}{E}\right)\right]\frac{2E(E-p_z)}{2E(E-p_z)+e^2A_c^2}\delta(p_{_\perp}-p'_{_\perp})
\delta(p_z-p'_z)\delta_{n,n'}\delta_{\sigma,\sigma'}\delta_{\tau,\tau'}\nonumber\\
&\!\!\!=&\!\!\!\frac{4\pi^2}{p_\perp}\delta(p_{_\perp}-p'_{_\perp})
\delta(p_z-p'_z)\delta_{n,n'}\delta_{\sigma,\sigma'}\delta_{\tau,\tau'}.\end{eqnarray}\end{widetext}

\end{document}